\documentclass[prd,aps,floatfix,superscriptaddress,nofootinbib,twocolumn,10pt,English]{revtex4-1}

\usepackage{amssymb,amsmath}
\usepackage{graphicx}
\usepackage[percent]{overpic}
\usepackage{overpic}
\usepackage{color}
\usepackage[colorlinks=true]{hyperref}
\usepackage{enumitem}
\usepackage{bm}

\begin{document}  

\title{Gravitational wave spectroscopy of binary neutron star merger remnants \\ with mode stacking} 
{
\author{Huan Yang}
\affiliation{Department of Physics, Princeton University, Princeton, New Jersey 08544, USA.}
\author{Vasileios Paschalidis}
\affiliation{Department of Physics, Princeton University, Princeton, New Jersey 08544, USA.}
\affiliation{Theoretical Astrophysics Program, Departments of Astronomy and Physics, University of Arizona, Tucson, AZ 85721, USA.}
\author{Kent Yagi}
\affiliation{Department of Physics, Princeton University, Princeton, New Jersey 08544, USA.}
\author{Luis Lehner}
\affiliation{Perimeter Institute for Theoretical Physics, Waterloo, Ontario N2L 2Y5, Canada}
\affiliation{CIFAR, Cosmology \& Gravity Program, Toronto, ON M5G 1Z8, Canada}
\author{Frans Pretorius}
\affiliation{Department of Physics, Princeton University, Princeton, New Jersey 08544, USA.}
\affiliation{CIFAR, Cosmology \& Gravity Program, Toronto, ON M5G 1Z8, Canada}
\author{Nicol\'as Yunes}
\affiliation{eXtreme Gravity Institute, Department of Physics, Montana State University, Bozeman, Montana 59717, USA}
\date{\today}

\begin{abstract} 

A binary neutron star coalescence event has recently been observed for
the first time in gravitational waves, and many more detections are
expected once current ground-based detectors begin operating at design
sensitivity.
As in the case of binary black holes, gravitational waves generated by
binary neutron stars consist of inspiral, merger, and post-merger
components.
Detecting the latter is important because it encodes information about
the nuclear equation of state in a regime that cannot be probed prior
to merger.
%
The post-merger signal, however, can only be expected to be measurable by current detectors
for events closer than roughly ten megaparsecs, which given merger rate
estimates implies a low probability of observation within the expected lifetime
of these detectors.
We carry out Monte-Carlo simulations showing that the dominant
post-merger signal (the $\ell=m=2$ mode) from individual binary
neutron star mergers may not have a good chance of observation even with the most
sensitive future ground-based gravitational-wave detectors proposed 
so far (the Einstein Telescope and
Cosmic Explorer, for certain equations of state, assuming a full year of operation, the latest merger
rates, and a detection threshold corresponding to a signal-to-noise
ratio of 5).
For this reason, we propose two methods that  stack the
post-merger signal from multiple binary neutron star observations to
boost the post-merger detection probability.
The first method follows a commonly-used practice of
multiplying the Bayes factors of individual events.
The second method relies on an assumption that the mode phase can 
be determined from the inspiral waveform, so that coherent mode stacking of the
data from different events becomes possible.
We find that both methods significantly improve the chances of
detecting the dominant post-merger signal, making a detection very
likely after a year of observation with Cosmic Explorer for certain
equations of state.
We also show that in terms of detection, coherent stacking is more
efficient in accumulating confidence for the presence of post-merger
oscillations in a signal than the first method.
Moreover, assuming the post-merger signal is detected with Cosmic
Explorer via stacking, we estimate through a Fisher analysis that the
peak frequency can be measured to a statistical error of $\sim$ 4--20
Hz for certain equations of state.
Such an error corresponds to a neutron star radius measurement to within
$\sim$ 15-56 m, a fractional relative error $\sim 4\%$, suggesting
that systematic errors from theoretical modeling $(\gtrsim$ 100 m) may
dominate the error budget.
 
\end{abstract}

\maketitle 

\section{Introduction}

The LIGO/Virgo collaboration recently announced the first detection of
a gravitational wave (GW) signal consistent with the inspiral
and merger of a binary neutron star (BNS) system~\cite{TheLIGOScientific:2017qsa},
corroborated by numerous observations of electromagnetic
counterparts across the spectrum, from radio to gamma rays~\cite{GBM:2017lvd}.
This one event has already provided a wealth of new information: highlights include
the establishment of a connection between NS mergers and (at least a class of) short gamma ray bursts,
evidence that a significant fraction of the universe's r-process elements
are born in NS mergers, an upper-bound constraint on the tidal deformability of neutron stars,
a measurement of the Hubble constant independent of the cosmic distance ladder, 
and a stringent constraint that the speed of gravitational waves equals the speed of light.

As loud as GW170817 was in gravitational waves with a network
signal-to-noise (SNR) ratio of 32, this still all came from the
inspiral phase of the event, and no detectable merger/post-merger
signal was reported by the LIGO/Virgo collaboration.  This is not
surprising, as regardless of what the outcome of the merger may have
been---prompt or delayed collapse to a black hole, or a stable high
mass NS remnant--- the corresponding GW emission is not expected to be
loud enough to allow extraction from the noise at the relevant
frequencies $> 1$kHz. Thus, at present only informed guesses as to the
nature of the remnant can be made, based on the consistency of models
of post-merger electromagnetic (EM) emission processes with
observations (insofar as the emission depends on properties of the
remnant; see e.g.~\cite{Metzger:2017wot}). Given how crude existing
models of the post-merger central engine of the EM counterparts are,
it would be ideal to instead measure properties of the remnant in GWs,
and use that to inform interpretation of the counterpart emission.

Beyond helping to decipher the EM data, the merger/post-merger GW signal
can contain much information of intrinsic value in understanding the physics
of the remnant. As mentioned, following the merger, the BNS
remnant may either promptly collapse to a black hole (BH), form a
supramassive neutron star or form a hypermassive neutron star (HMNS)
that will ultimately undergo delayed collapse to a BH
(see~\cite{Paschalidis:2016agf} for a recent review). The latter two
scenarios lead to a remnant that spins rapidly and undergoes
non-axisymmetric oscillations, emitting GWs in the process. More than
a decade of simulations of BNS mergers have revealed that the
post-merger GW spectrum is rich, with several distinct peaks that can
be used to probe the merger remnant through spectroscopy (see
e.g. \cite{Oechslin2002PhRvD..65j3005O,ShibataUryu2002,Shibata2005PhRvD..71h4021S,Shibata:2006nm,Kiuchi2009PhRvD..80f4037K,Bauswein:2011tp,Takami2014,bauswein2015exploring,Palenzuela:2015dqa,Lehner:2016lxy,Rezzolla:2016nxn,Dietrich:2016lyp,Radice:2016rys}
for some historical and recent work,
\cite{PhysRevD.86.104006,PhysRevD.88.044047,PhysRevD.87.041502,PhysRevLett.114.081101}
for related work on BH spectroscopy,
and~\cite{Baiotti:2016qnr,Paschalidis:2016vmz} for recent reviews). In
the first 10-20 ms after merger, the dominant component of a
post-merger GW is the $\ell=m=2$ mode (which we call here the 22 mode
for short). For BNS merger remnants that may survive for longer times,
a one-arm mode ($\ell=2,m=1$, or 21 mode) can dominate the GW
emission~\cite{Paschalidis:2015mla,East:2015vix,East:2016zvv,Lehner:2016wjg,Radice:2016gym}.

Extracting post-merger information from GWs is also crucial for obtaining
a full understanding of the physics of nuclear matter.
Individual NSs (with mass $M_{\rm NS}$) in inspiralling
binaries are described by cold nuclear matter, whereas BNS merger
remnants (with mass $\sim 2M_{\rm NS}$) are described by hot nuclear
matter. Therefore, GWs from BNS merger remnants encode the physics of
dense nuclear matter in a regime that is not accessible in the inspiral
phase. In addition, measured post-merger GWs could reduce the uncertainties in 
information drawn from the inspiral phase, just as with
binary BH mergers
(e.g.~\cite{TheLIGOScientific:2016wfe,TheLIGOScientific:2016src}).
Moreover, these waves will provide further insight to help disentangle
degeneracies between modulations due to tidal effects from
those induced by deviations from General Relativity
(e.g.~\cite{Sampson:2014qqa,ST1,Shibata:2013pra,Palenzuela:2013hsa}).}
We can also anticipate that information on the interior composition of cold neutron stars will
be available through independent electromagnetic observations, for example with the
recently launched NICER~\cite{doi:10.1117/12.2231304} (under suitable
assumptions, NICER may determine neutron star radii to $\simeq 5\%$
accuracy which, in turn, will help constrain the cold nuclear
EOS). One can then envision either employing such knowledge to further
constrain GW predictions, or use the GW observations independently 
and crosscheck for consistency with results from EM observations.

GWs from BNS post-merger oscillations are challenging to detect, as
indicated by previous studies~\cite{Clark:2014wua,clark2016observing},
and evidenced by a lack of detection of any with GW170817 \cite{abbott2017search}. With
certain binary parameters, EOS, etc., an event not too much closer
than GW170817 could produce a detectable post-merger signal with aLIGO
sensitivities. However, even with the more optimistic BNS merger rates
of $1540 (+3200 -1220) /{\rm Gpc}^3/{\rm yr}$ implied by
GW170817~\cite{TheLIGOScientific:2017qsa}, a similarly loud merger is
roughly a once-per-decade event.  On the other hand, these merger
rates suggest several events (including GW170817) could be expected in
the aLIGO era, and even more in the era of third-generation
ground-based detectors, that are all within a factor of a few in SNR of
having individually detectable post-merger signals.
If there is a common post-merger signal in these anticipated events,
we can therefore attempt to go after the common component by combining,
or stacking, the data from multiple events appropriately. The data may also be combined 
through unmodelled algorithms (e.g., \cite{chatziioannou2017inferring}) for parameter estimation purpose.

To simplify the analysis in our first study of this idea, 
we only include the 22 mode of the post-merger
signal. We model it as an exponentially decaying sinusoidal
function, which is consistent with the leading order behavior
identified in the principal component analysis of 
BNS post-merger waveforms in \cite{clark2016observing}.
We propose two methods to
stack this data from different detections. With the first method,
we treat all events as independent and combine the 
Bayes factor, following a similar approach as discussed in \cite{meidam2014testing} (referred to as ``power stacking" in this work). 
In the second method, we assume that the theoretical uncertainties in future numerical BNS simulations
can be significantly reduced, such that the inspiral waveform can be used to predict 
the phase of post-merger modes. In this case, the dominant modes from different post-merger signals 
can be coherently stacked together,
as shown in the black hole ringdown scenario \cite{yang2017black}.
Essentially, coherent mode stacking is the shifting
and rescaling of $N$ signals to align their phase 
using information from the inspiral in
order to construct a weighted, linear superposition that boosts the
post-merger SNR. Both methods are able to boost the detectability
of post-merger oscillations. The coherent stacking approach outperforms
the first method by taking advantage of the additional phase information.


There are several important issues to note in this work. 
First, the coherent stacking procedure presented here is similar to the treatment we
developed in \cite{yang2017black}, which was designed to boost
particularly relevant features in signals from binary black hole
mergers, for instance the SNR of secondary modes in BH
ringdowns. However, there are important differences between the
stacking approach developed in this paper and in \cite{yang2017black}.
In particular, the inspiral-merger-ringdown waveform of binary BHs is known
from numerical relativity simulations sufficiently accurately that it
can be used to predict the phase of secondary modes, which in turn set
the basis to align the secondary modes from different detected
events. By contrast, current numerical relativity simulations of BNS
mergers cannot reliably determine the phase of post-merger
oscillations, partly because there are important pieces of physics
(such as turbulent magnetohydrodynamics, microphysical effects, NS
spin effects etc.)  that are not fully resolved or accounted
for. Despite the significant progress in our understanding of BNS
post-merger physics (see~\cite{Baiotti:2016qnr,Paschalidis:2016vmz}
for recent reviews), there remain obstacles both in the computational
aspect and the physical understanding of the problem that must be
overcome before reliable GWs from numerical relativity simulations can
be used to construct GW templates. Therefore, in this work we
emphasize the application of power stacking, and also generalize the hypothesis test formalism (Generalized Likelihood Ratio
Test or GLRT) of~\cite{yang2017black} to signals with unknown phase.

Second, although for a given EOS the frequency (unlike the phase) of
the $22$ mode can be robustly determined by numerical
simulations~\cite{bauswein2015exploring}, the true underlying EOS is
unknown. Thus, in order to perform the hypothesis test for detection, we
assume an underlying EOS to compute the mode frequencies for each
event. Picking an incorrect EOS would in principle generate frequency
mismatch which would
degrade the SNR of the signal. On the other hand, one can perform a model-selection study to
compare different EOSs for their relative consistency with the
data. We investigate this issue here as well.

Third, as mentioned earlier, the BNS merger remnant can undergo
collapse to a BH promptly after the merger, in which case there is no
GW signal from a HMNS to stack. According to the work
of~\cite{Shibata:2006nm} there exists a threshold total binary mass
that determines whether prompt collapse will take place, independently
of the mass ratio. Therefore, we only consider events with total mass
below this threshold in our Monte-Carlo (MC) simulations before
stacking. Since in this study we focus on finite-temperature,
realistic nuclear EOSs we use the threshold masses for prompt collapse
determined in~\cite{bauswein2013prompt}. 
We show how detection/non-detection of a stacked signal
from a suitable population of events can provide a direct test of this
collapse hypothesis, and further be used to place constraints on the nuclear EOS.

Finally, the starting time of coalescence may be subject to systematic uncertainties in modeling the tidal
effects of binary NSs in the inspiral stage using post-Newtonian methods. This does not
significantly affect the calculations in this paper, as we mainly focus
on the properties of the $22$ (peak) mode, which radiates waves with
frequency above $2 \rm kHz$ (well above the merger
frequency). In addition, an accurate numerical waveform would naturally
take into account all tidal effects.

\subsection{Executive summary}

We now summarize the main results of this paper. Unless otherwise
specified, for the sake of presentation our calculations will focus on
the Cosmic Explorer (CE) experiment as the representative third-generation instrument;
we expect similar conclusions to hold for both CE and the Einstein Telescope (ET).  Based on
the MC simulations we have performed, given an EOS
(TM1~\cite{Hempel:2011mk} for reference) and with the adopted BNS
merger rate, the chances of detecting a single post-merger event after
 one year of observations with third
generation detectors are good, but not certain (for simplicity, here and henceforth we
use the word ``event'' to only refer to the post-merger signal).
By stacking the loudest events, a detection becomes almost certain after a
year of observations with CE. For example, if the SNR threshold for
detection is set to 5, a one-year observation with CE  has a
$\sim 79\%$ chance of detecting a post-merger oscillation signal in a
single event, while the chances increase to $\sim 100\%$ after power
stacking the top 5 loudest events.

Apart from power stacking that
simply multiplies the Bayes factor of each
event~\cite{Agathos:2013upa,PhysRevD.90.064009,Agathos:2015uaa}
\footnote{Calling this ``power stacking'' is
  a slight abuse of historic notation, as this term has mostly been
  used to refer to analysis strategies that add excess power in select
  tiles in a time-frequency decomposition of multiple signals; see
  e.g. ~\cite{Kalmus:2009uk,Tai:2014bfa}. These methods also give a
  composite SNR that scales as $N^{1/4}$ for N identical events each
  with low individual SNR, as multiplication of Bayes factors
  does~\cite{yang2017black}, which is why we have borrowed this
  nomenclature.}, we also investigate combining signals if the phase of the post-merger 
modes can be predicted using simulations informed by source parameters measured
from the inspiral waveform. This is reasonable to expect by the era of third generation
gravitational wave observatories, as future numerical modelling
of binary neutron star mergers is anticipated to become sufficiently
accurate by then. We compare these two methods and find that coherent stacking
is more efficient at enhancing the SNR of BNS post-merger signals than
power stacking (see Sec.~\ref{sec32} and Fig.~\ref{fig:alpha}).  This
is partially because the coherent stacking method we propose requires
extra phase information.

We also carry out a Bayesian model selection analysis to see how well
one can distinguish between two different EOS models. For example, the
TM1 EOS can be well distinguished from the DD2
EOS~\cite{Hempel:2011mk}, with the average log-Bayes factor in the
range $20$--$100$ using the single loudest event (and $130-300$ for power-stacked signals).  
We further perform a parameter
estimation study to derive how accurately one can measure the peak
frequency of post-merger oscillations. We convert such a statistical
error on the peak frequency to a statistical error on the NS radius of
a $1.6M_\odot$ NS using a universal relation between these
quantities~\cite{bauswein2015exploring}.  We find that with the
power-stacked  signal and using CE, the statistical error on the NS radius
ranges from 15 m to 56 m, depending on the underlying EOS, which
constitutes a fractional relative error of $\sim 4\%$. Such a
measurement would thus compete with NICER measurement of the
mass-radius relation of isolated NSs~\cite{2012SPIE.8443E..13G}.
However, at this time systematic error in the universal relations
between post-merger oscillation frequency and binary total mass, as
well as in the template construction, dominate over the statistical
error; this may be reduced in the future through, e.g., better
modeling of NSs and more accurate BNS merger simulations.

A recent study by Bose et al. \cite{bose2017neutron} also proposes to
perform stacking of multiple BNS post-merger events with a focus on
parameter estimation. Apart from considering only second-generation
gravitational-wave detectors, and using results from simulations that
do not employ finite temperature EOSs, our results are distinct from
this work in at least two additional, significant aspects. First, we
focus on the dominant $22$ mode modeled as a single damped sinusoid,
while in \cite{bose2017neutron} a several parameter fit to the entire
post-merger signal is used, with additional assumptions on the phases
of the modes used in the fit. This alters the parameter estimation
because the target templates are different. Second, here we perform a
more detailed investigation and in-depth study to assess the
detectability of GWs from BNS merger remnants by applying a
Generalized-Likelihood-Ratio-Test (for hypothesis testing). In
particular, instead of focusing on parameter estimation alone, we also
discuss in detail the performance of stacking methods in making a
detection of the dominant $22$ mode, as well as a comparison to the
power stacking method. This discussion is necessary because strictly
speaking the interpretation of results of parameter estimation using
modes from a set of events is only valid after a statistically significant
confirmation of the existence of the modes have been made.

\subsection{Organization}

This paper is organized as follows. In Sec.~\ref{sec2} we develop the
hypothesis test (GLRT) formalism for signals with unknown phase and
perform a MC study to probe the detectability of post-merger
oscillations from individual BNS remnants, assuming several different
EOSs\footnote{Throughout this work we assume that in nature neutron
  stars have a unique EOS. This is a standard assumption, though one
  could envision mass-dependent EOS variations, or more unusual
  situations where for example strange quark stars and conventional
  neutron stars can both exist in the same mass range.} and focusing
on third-generation GW detectors.  In Sec.~\ref{sec3} we apply the
hypothesis test (GLRT) formalism for signals with unknown phase to
stacked signals. We use the individual signals from the MC study of
Sec.~\ref{sec2} to demonstrate that these stacking methods
significantly amplify the SNR of BNS post-merger GWs and their
detectability. Moreover, we show that coherent stacking works more
efficiently than power stacking.  In Sec.~\ref{sec4} we discuss the
possibility of distinguishing different EOSs using the stacked signal
by carrying out a Bayesian model selection study. We also perform a
parameter estimation study to derive the measurement accuracy of the
post-merger peak frequency, and in turn, that of the NS radius.  We
conclude in Sec.~\ref{sec5} and discuss possible directions for future
work.

\section{Single event detection}\label{sec2}

In this section we present the GLRT formalism we develop for single
events, and perform a MC study to assess the detectability of
BNS post-merger oscillations from individual events using
third-generation ground based GW interferometers.

\subsection{Hypothesis testing with unknown phase} \label{sec21}

Let us begin by describing how we generalize the Bayesian hypothesis test
formalism of \cite{yang2017black,Berti:2007zu,Berti:2016lat} such that
it is applicable to coherent stacking of signals without prior phase
information. In this section, we extend the formalism to the case of individual
signals with unknown phase offset, which is suitable for finding
oscillations of BNS merger remnants. In Sec.~\ref{sec31}, we describe the procedure of
coherently stacking a set of events, which involves frequency
rescaling and phase alignment.

As we mentioned in the introduction, in the first $10-20$ ms following
a typical BNS merger the 22 mode is the dominant one. Thus, instead of
trying to model full signatures of post-merger waveforms, we focus on
the dominant peak of the 22 mode component (see also the Principal
Component Template in \cite{clark2016observing}).  Here, we model the
$22$ mode oscillation as a damped sinusoid
\begin{align}\label{eq:w22}
h(t) = A'A_r \sin(2\pi
f_{\rm peak} t- \phi^0) e^{- \pi f_{\rm peak} t/Q} \Theta(t)\,,
\end{align}
where $A'$ is the amplitude, $f_{\rm peak}$ is the $22$ mode peak
oscillation frequency, while we label the time coordinate in a way
that the waveform starts at $t=0$ (hence the Heaviside step function
$\Theta(t)$) and $\phi^0$ is a constant phase offset. The factor $A_r$
denotes the reduction of the wave amplitude arising from source
inclination and the response of the detector. Finally, $Q$ is the quality
factor of the mode.

We now explain the GLRT formalism and its extension. The one used 
in~\cite{yang2017black,Berti:2007zu,Berti:2016lat} assumes that 
all the parameters in the waveform are known \emph{a priori} except for the amplitude. 
In our context, we assume that $f_{\rm peak}$, $Q$ and $A_r$ are known from
the inspiral information together with a given underlying EOS.
One can then repeat the analysis with a different choice of EOS and carry out
a Bayesian model selection study to see which one is preferred (see Sec.~\ref{sec4a}).
On the other hand, the phase
$\phi^0$ is unknown for BNS post-merger GWs, which requires one to extend 
the GLRT formalism.
We begin by rewriting 
Eq.~\eqref{eq:w22} as
\begin{align}\label{eq:w22d}
h(t) = &\left [ A_s' \sin(2\pi f_{\rm peak} t) + A_c' \cos(2\pi f_{\rm peak} t)\right ] e^{- \frac{\pi f_{\rm peak} t}{Q}} \Theta(t)\,\nonumber \\
= & A_s h_s(t)+A_c h_c(t)\,,
\end{align}
with $A'A_r=\sqrt{A_c'^2+A_s'^2}$ and $\tan \phi^0
=-A_c'/A_s'$. Here, $h_c$ ($h_s$) is proportional to the above cosine
(sine) function with an arbitrary normalization constant.  Therefore,
testing for a signal with unknown phase can be phrased as a test between 
the following two hypotheses:
 \begin{align}
 \mathcal{H}_1: \,\, & \tilde y(f) =A_c \tilde h_c(f) +A_s \tilde h_s (f) +\tilde n(f)\,, \label{hypothesis1}\\
 \mathcal{H}_2: \,\, & \tilde y(f) = \tilde n(f) \,,
  \end{align}
with $A_c^2+A^2_s>0$ in $\mathcal{H}_1$. Here $\tilde h_c$ and $\tilde h_s$ are two
frequency-domain bases of the waveform which are
nearly orthogonal to each other (this is generally true if $Q \gg 1$, i.e. there are enough
cycles in the relevant waveform),
so that $\langle h_c |h_s \rangle \approx 0$, where the inner product is
defined as
\begin{align}
 \langle \chi | \xi\rangle \equiv 2 \int^\infty_0 \frac{\tilde \chi^*(f) \tilde \xi(f) +\tilde \chi(f) \tilde \xi^*(f)}{{S}_n} df\,,\quad ||\xi||^2 \equiv \langle \xi | \xi\rangle
 \end{align}
with respect to the one-sided spectral density of detector noise
$\langle \tilde n(f) \tilde n^*(f') \rangle =[S_n(f)/2] \,\delta (f-f')$.  

The posterior probability of a hypothesis $\mathcal H$ being correct given some data $y$ 
is given by Bayes' theorem~\cite{Cornish:2007if,Littenberg:2009bm}
\begin{equation}
P(\mathcal{H}|y) = \frac{P(\mathcal{H})\, P(y|\mathcal H)}{P(y)}\,,
\end{equation}
where $P(\mathcal{H})$ is the prior belief in $\mathcal H$, while $P(y)$ 
is the probability of the data, which serves as an irrelevant normalization constant.
The evidence $P(y|\mathcal H)$ is given by
\begin{equation}
\label{eq:evidence}
P(y|\mathcal H) \equiv \int d \vartheta\,P(\vartheta|\mathcal H) \, P(y| \vartheta \mathcal H)\,,
\end{equation}
where $P(\vartheta|\mathcal H)$ is the prior on the model parameters $\vartheta$, while 
$P(y| \vartheta \mathcal H)$ is the likelihood function.

The likelihood of Hypothesis 1 is given by
\begin{align}\label{eqprob}
 P(y|\vartheta^{i} \mathcal{H}_1) &\propto \prod_{f>0} {\rm exp} \left ( -\frac{2 |\tilde y-A_s \tilde h_s-A_c \tilde h_c|^2}{S_n}\right )\,\nonumber \\
 & \propto {\rm exp}\left (- \frac{||y-A_s h_s-A_c h_c||^2}{2}\right )\,,
\end{align}
where $\vartheta^{i}=\{A_c,A_s\}$. For a uniform prior on $A_c$
and $A_s$, the marginalization over $\vartheta$ in
Eq.~\eqref{eq:evidence} corresponds to maximizing the above likelihood
over $A_c$ and $A_s$. The maximum likelihood estimator, using 
the shorthand notation  $c=h_c$ and $s=h_s$, is then given
by
\begin{align}\label{eqmaxl}
& \hat{A}_c =\frac{\langle c | y \rangle}{\langle c | c\rangle}\,, \quad \hat{A}_s =\frac{\langle s | y \rangle}{\langle s | s \rangle}\,.
\end{align}
Thus, according to Eqs.~\eqref{eq:evidence},
\eqref{eqprob} and the discussion above Eq.~\eqref{eqmaxl}:
\begin{align}\label{PyH1}
P(y|\mathcal H_1) \propto {\rm exp}\left (- \frac{||y-\hat A_s s- \hat A_c c||^2}{2}\right )\,,
\end{align}
and consequently
\begin{align}
P(\mathcal H_1|y) &\propto P(\mathcal{H}_1)\, P(y|\mathcal H_1) \nonumber \\
&\propto P(\mathcal{H}_1)\,{\rm exp}\left (- \frac{||y-\hat A_s s- \hat A_c c||^2}{2}\right )\,.
\end{align}
Repeating these steps for Hypothesis 2 ($A_s=A_c=0$) gives
\begin{equation}\label{PyH2}  
  P(y|\mathcal{H}_2) \propto {\rm exp}\left (- \frac{||y||^2}{2}\right )\,,
\end{equation}
\begin{equation}
P(\mathcal H_2|y) \propto P(\mathcal{H}_2)\,{\rm exp}\left (- \frac{||y||^2}{2}\right )\,.
\end{equation}

The betting odds
of $\mathcal H_1$ over $\mathcal H_2$, known as the odds ratio, is given by 
\begin{align}\label{eqodd}
O_{12} \equiv \frac{P( \mathcal{H}_1 | { y})}{P( \mathcal{H}_2 | { y})} = \frac{P(\mathcal{H}_1)}{P(\mathcal{H}_2)} B_{12}\,,
\end{align}
where 
\begin{align}
\label{eq:BF}
B_{12} \equiv \frac{P(y|\mathcal H_1)}{P(y|\mathcal H_2)}
\end{align}
is the Bayes factor. We focus on using $B_{12}$ throughout this paper,
though it agrees with $O_{12}$ in the case of equal priors
$P(\mathcal{H}_1) = P(\mathcal{H}_2)$ or in the case of uninformative
priors $P(\mathcal{H}_1) = 0.5$ and $P(\mathcal{H}_2) = 0.5$.

We next compute the log of the Bayes factor, which using Eqs.~\eqref{eqmaxl}, \eqref{PyH1} and~\eqref{PyH2} is

\begin{align}
\label{eqso}
\hat{T}_{\rm single} & \equiv \log \left [ \frac{P(y|\mathcal H_1)}{P(y|\mathcal H_2)} \right ]_{A_{c,s} \rightarrow \hat{A}_{c,s} } 
=\frac{\langle c | y \rangle^2}{ 2\langle c | c \rangle}+\frac{\langle s | y \rangle^2}{ 2 \langle s | s \rangle}\, \nonumber \\
& =\frac{\langle c| c \rangle}{2} (A^2_c+A^2_s)\\
& +\frac{\langle c | n \rangle^2+\langle s | n \rangle^2}{2\langle c | c \rangle }+\frac{A_s \langle s | n \rangle+A_c \langle c | n \rangle}{\langle c | c \rangle} \nonumber \\
&=s_T+n_T\,,
\end{align}
where $s_{T}$ is defined as the first term in equation (\ref{eqso})
i.e.,~$s_{T}={\langle c| c \rangle} (A^2_c+A^2_s)/2 \, (=\rho^2/2)$, and
the remaining terms are defined as $n_{T}$.  Note that in going from
the first line to the second and third lines in the above we
replaced the data $y$ on the right-hand-side of the equality in the
first line by Eq.~\eqref{hypothesis1}. Here, we have chosen the
normalization\ of $h_c$ and $h_s$ such that $\langle c|c\rangle=
\langle s|s \rangle$. If we are to take the log of the odds ratio 
(defined in Eq.~\eqref{eqodd}) instead of the log of the Bayes factor
and $P(\mathcal{H}_1) \neq P(\mathcal{H}_2)$,
$s_T$ needs to be shifted by $\log
        [{P(\mathcal{H}_1)}/{P(\mathcal{H}_2)}]$.

The evidence to favor (or disfavor) $\mathcal{H}_1$ over
$\mathcal{H}_2$ depends on the signal part $s_T$, and the distribution
of the noise part $n_T$.  The GLRT ratio variable $\hat{T}_{\rm
  single}$ can be intuitively thought of as an approximate spectral
power of $y$ near the central frequency $f_{\rm peak}$. The
distribution of $n_T$ is in general non-Gaussian, but when
$A_{c,s}=0$, it becomes $\chi^2_2$ (chi-squared with 2 degrees of
freedom). Here and throughout we assume that $\langle c|n\rangle$ and
$\langle s|n\rangle$ are normally distributed.
If we denote the right-tail probability function of $n_T$ 
(whose probability distribution is $P_{n_T}$) as
\begin{equation}
\label{eq:right-tail}
R(x) = \int^\infty_x P_{n_T}(z) dz\,,
\end{equation} 
(with $R_{A_{c,s}=0}$ corresponding to that of the $\chi^2_2$
distribution) and the false-alarm probability is $P_f$, the criteria
for rejecting hypothesis $\mathcal{H}_2$ with $\hat{T}_{\rm single}$
computed from observation data is
\begin{align}
R_{A_{c,s}=0} (\hat{T}_{\rm single}) \le P_f\,,
\end{align}
or
\begin{align}
\hat{T}_{\rm single}  \ge R^{-1}_{A_{c,s}=0} (P_f)\,.
\end{align}
Now, notice that under $\mathcal{H}_1$, $\hat{T}_{\rm single}=s_T+n_T$ is a random variable depending on the underlying signal and detector noise. Based on its distribution, one can infer the probability that the above inequality is satisfied, giving the target detection rate (probability) $P_d$:
\begin{align}
P_d 
\ge R(R^{-1}_{A_{c,s}=0} (P_f)-s_T)\,.
\end{align}
The amplitude of signal $A_{c,s}$ 
required to satisfy the above bound is then

\begin{align}\label{eq:pass}
\frac{1}{2}\langle c| c \rangle (A^2_c+A^2_s) \ge R^{-1}_{A_{c,s}=0} (P_f)-R^{-1} (P_d)\,,
\end{align} 
or equivalently, the SNR
required to satisfy the bound is given by
\begin{align}
\rho &\ge \rho_\mathrm{thres} \equiv \sqrt{2 [R^{-1}_{A_{c,s}=0} (P_f)-R^{-1} (P_d)]}\,,
\end{align}
where we used the relation $s_T = \rho^2/2$.

The probability distribution function $P_{n_T}(z)$ inside the integral of 
the right-tail probability function $R$ in Eq.~\eqref{eq:right-tail}
is obtained as follows. For simplicity,
we choose the normalization such that $\langle c | c \rangle =1= \langle
s | s \rangle $, and denote $X \equiv \langle c | n \rangle^2/2 +
A_c \langle c | n \rangle$, $Y \equiv \langle s | n \rangle^2/2 + A_s
\langle s | n \rangle$. Given that $X$ and $Y$ are independent
  random variables, the probability distribution of the random
  variable $Z=X+Y$ is given by
\begin{align}\label{eqcomd0}
P_{Z} (z) = \int^{\infty}_{-\infty} P_X(x') P_Y(z-x') d x'\,.
\end{align}
However, notice that by definition $X = (\langle c | n
  \rangle+A_c)^2/2 -A^2_c/2 \ge -A^2_c/2$, and similarly $Y \ge
  -A^2_s/2$. Therefore, Eq.~\eqref{eqcomd0} becomes
\begin{align}\label{eqcomd}
P_{n_T} (z) = \int^{z+A^2_s/2}_{-A^2_c/2} P_X(x') P_Y(z-x') d x'\,.
\end{align}
Here
\begin{align}
P_X(x') = \sqrt{\frac{2}{\pi}}\frac{e^{-A^2_c-x'} \cosh [A_c \sqrt{A^2_c+2 x'}]}{\sqrt{A^2_c+2 x'}}\,,
\end{align}
which is obtained from a non-central $\chi^2$ distribution with an appropriate change of variable. 
In fact, $P_{n_T}$ can also be obtained from a non-central $\chi^2_2$ distribution with an appropriate change of variable.
As expected, $P_X$ reduces to a Gaussian distribution in the large $A_c$ limit, and reduces to 
the $\chi^2$ distribution with one degree of freedom for $A_c \to 0$. 
The distribution $P_Y $ follows similarly, with $c \rightarrow s$. 
For completeness, we show the variance of $n_T$ in Appendix~\ref{appendixA}. 
We also show 
the signal-to-noise level of $\hat T$ that one can use instead of $\rho$ to discuss the 
detection criterion.

There are two important facts regarding $R$. First, the distribution $R$ depends on $A_{c,s}$ only through $A^2_c+A^2_s$. This can be seen by writing  $n_T$ as 
\begin{align}
n_T =& A_s \langle s | n\rangle +A_c \langle c |n \rangle
+\frac{1}{2(A^2_c+A^2_s)} (A_c \langle c | n\rangle +A_s \langle s |n \rangle )^2 \nonumber \\
&+\frac{1}{2(A^2_s+A^2_s)} ( A_s \langle c | n\rangle -A_c \langle s |n \rangle )^2\,,
\end{align}
and noting that $A_c \langle c | n\rangle +A_s \langle s |n \rangle$ and
$A_s \langle c | n\rangle -A_c \langle s |n \rangle$ are independent
Gaussian random variables with the same variance, $A^2_c+A^2_s$.
Second, if $A_{c,s}=A^m_{c,s}$ is the marginal solution that satisfies
the equality in Eq.~\eqref{eq:pass}, and if we further scale the
detector noise such that $n \rightarrow C n$ without changing the
definition of the inner product $\langle | \rangle$ so that $c$ and $s$ do not
have to be renormalized, it is straightforward to see that $A_{c,s} =C
A^m_{c,s}$ still satisfy the equality in Eq.~\eqref{eq:pass} with the
rescaled noise. Such a property is important as it means there is a
one-to-one mapping between the threshold event SNR (schematically $\sim \sqrt{A_c^2+A_s^2}/n$) and $P_f$
and $P_d$. Such a property also carries over to the stacked signal we
consider in Sec.~\ref{sec31}. Following the convention in
\cite{clark2016observing}, we shall set the threshold SNR to
$\rho_{\rm thres} = 5$, which is consistent with setting $P_f=0.01$ and
$P_d =0.982$.

\subsection{MC study}\label{sec22}

The detectability of post-merger oscillations from BNS remnants is
discussed in \cite{clark2016observing}, assuming optimal sky
orientation and source inclination. The results indicate that
post-merger oscillations from individual sources are detectable only
by third-generation GW detectors. Here we extend the analysis, but
with two important modifications that make the analysis more realistic
(although unfortunately greatly reducing detectability):

\begin{itemize}

\item[1.] Instead of assuming the optimal sky location and source inclination that
maximize the SNR, we randomly sample sources in sky direction, orbit
inclination angle and polarization angle. According to
\cite{sathyaprakash2009physics,regimbau2012mock}, the sky-averaged
amplitude for a given type of source receives a $2/5$ reduction
factor compared to the optimized configuration (assuming an
``L''-shaped GW detector). In addition, the opening angle between arm
cavities in the design for ET is $60^\circ$, leading
to an overall $\sqrt{3}/2$ reduction in signal amplitude comparing to
an ``L''-shaped interferometer with the same arm length.  In our
MC simulations, we have a different reduction factor for each
source based on its parameters, although on average it recovers the
$2/5$ factor for ``L''-shaped detectors. To obtain the reduction
factor for each source, we assume the ``L"-shape antenna pattern
function \cite{sathyaprakash2009physics} for CE:
\begin{align}
F_+ &= \frac{1}{2} (1+\cos^2\theta) \cos 2\phi\, \cos 2 \psi \nonumber \\
& -\cos \theta\, \sin 2\phi\,\sin 2\psi \,, \\
F_\times &= \frac{1}{2} (1+\cos^2\theta) \cos 2\phi\, \sin 2 \psi \nonumber \\
&+\cos \theta\, \sin 2\phi\,\cos 2\psi\,,
\end{align} 
and the single-detector antenna pattern function for ET \cite{regimbau2012mock}:
\begin{align}
 F_+ & = -\frac{\sqrt{3}}{4} \left [ (1+\cos^2\theta) \sin 2\phi\, \cos 2 \psi \right. \nonumber \\
 & \left. +2\cos \theta\, \cos 2\phi\,\sin 2\psi \right ] \,, \\
 F_\times & = \frac{\sqrt{3}}{4} \left [ (1+\cos^2\theta) \sin 2\phi\, \sin 2 \psi \right. \nonumber \\
 & \left.  -2\cos \theta\, \cos 2\phi\,\cos 2\psi \right ]\,.
\end{align} 
Here $\theta$ and $\phi$ are the angular coordinates of the source in the
detector frame and $\psi$ is the polarization angle. The amplitude
fraction in each polarization can be computed by
\begin{align}
\label{eq:ApAc}
\mathcal A_+ = F_+ \frac{1+\cos^2\varsigma}{2}\,,\quad \mathcal A_\times = F_\times \cos \varsigma\,,
\end{align}
where $\varsigma$ is the inclination angle of the BNS orbit with
respect to the line of sight. The
overall amplitude reduction factor with respect to the optimal
configuration that enters in Eq.~\eqref{eq:w22} is then given by
$A_r=\sqrt{\mathcal A^2_++\mathcal A^2_\times}$ for an ``L"-shape
detector and $A_r=(2/\sqrt{3}) \sqrt{\mathcal A^2_++\mathcal
  A^2_\times}$ for a single detector following the ET design. If we
allow three detectors placed in a triangle geometry as explained in
\cite{regimbau2012mock}, the corresponding factor is
$A_r=2/\sqrt{3}\sqrt{\sum_{i,j} \mathcal A^2_i
  (\theta,\phi+2\pi j/3,\psi,\varsigma)}$ 
  with the summation over 
  $i=(+,\times)$ and $j=(-1,0,1)$. The total SNR receives a
factor of $\sqrt{3}$ boost on average compared to the single detector case. 
In fact, if there are $N_d$
identical detectors, the total SNR is a factor of $\sqrt{N_d}$ larger
than the single detector SNR.

\item[2.] We adopt a more up-to-date estimate of the BNS merger rate
  from~\cite{TheLIGOScientific:2017qsa} based on the observed BNS merger. Such a rate ($R_{BNS} =
  1.54 \; {\rm Mpc}^{-3} \; {\rm Myr}^{-1}$) is
  $1.5$ times higher than
  the ``realistic" rate of \cite{abadie2010predictions}, which was
  adopted in \cite{clark2016observing}. Naturally, as a result, we
  predict more detections over a one-year observation period. Our
  conclusions can be easily modified if the true rate turns out to be
  different than this number. An argument about the relevant scaling
  goes as follows. Considering the case where for a given rate
  $R_{BNS}$ only one event is above the detection threshold within a
  volume of space $V$ after $T_{\rm obs}= 1$ yr of observations, we
  have $R_{BNS}\times V\times T_{\rm obs}=1$. But, $V \propto d^3
  \propto 1/\rho^3$, with $d$ the distance. Thus, the SNR should scale
  with the merger rate as $\rho \propto R_{BNS}^{1/3}T_{\rm
    obs}^{1/3}$. In reality different merger events are not identical,
  their source parameters and sky locations all affect their SNR, and
  for sufficiently high redshift $V$ is not simply proportional to
  $d^3$. Nevertheless, the above simple expression can be used to
  approximately scale the SNR that we present in our study below for
  different merger rates or observation periods.

\end{itemize}

Our analysis is based on the waveform model of Eq.~\eqref{eq:w22}, which
depends on the
peak frequency $f_{\rm peak}$, the quality factor $Q$ and the 22 mode amplitude
$A'$, the angle-dependent amplitude factor $A_r$ and phase offset $\phi^{0}$. 
We estimate the $22$ mode frequency ($f_{\rm peak}$) using the
fit of \cite{bauswein2015exploring} (see
also~\cite{Bauswein:2012ya,Bauswein:2014qla,Takami2014,Takami:2014tva,Bauswein:2015yca,Lehner:2016lxy}
for other fits)
\begin{align}\label{eq:fp}
\frac{f_{\rm peak}}{1  {\rm kHz}} 
= \frac{m_1+m_2}{M_\odot} \left [a_2 \left(\frac{R_{1.6 M_\odot}}{1 {\rm km}} \right)^2 +a_1 \frac{R_{1.6 M_\odot}}{1 {\rm km}} +a_0\right ]\,,
\end{align}
where $a_0=5.503$, $a_1=-0.5495$ and $a_2=0.0157$ are EOS-independent parameters;
$R_{1.6 M_\odot}$ is the radius of a non-rotating NS with gravitational mass
$1.6M_\odot$, and this parameter therefore encodes the EOS dependence.
We choose the masses by 
independently sampling the Gaussian distribution 
\cite{ozel2016masses}
\begin{align}
P(M_{\rm NS};M_0,\sigma) = \frac{1}{\sqrt{2\pi \sigma^2}}\exp\bigg[-\frac{(M_{\rm NS}-M_0)^2}{2\sigma^2}\bigg] \label{Mnsdistribution}
\end{align}
with $M_0=1.33M_\odot$ and $\sigma=0.09M_\odot$.

The quality factor and $22$ mode amplitude 
in Eq.~\eqref{eq:w22} should also depend on the NS EOS, the mass ratio
and mass of the binary, but the detailed dependence is currently
unknown. In order to enable comparison to the results in
\cite{clark2016observing}, we set $A'$ and $Q$ such that the peak
value of the characteristic strain and the SNR of Eq.~\eqref{eq:w22}
match the peak characteristic strain and SNR of the dominant $22$ mode
component in Fig.~$11$ of \cite{clark2016observing}, which corresponds
to the post-merger signal arising from a $1.35 M_\odot+1.35 M_\odot$
BNS with optimal extrinsic parameters (sky location and inclination
angle), at luminosity distance $d= 50 {\rm Mpc}$ with the Hempel et
al. EOS (TM1)~\cite{Hempel:2011mk}. The matching process yields
$Q=34$, $A'=2.5 \times 10^{-22}$. For a binary obeying the TM1 EOS,
but with different component masses and luminosity distance we still
set $A'$ based on a $1.35 M_\odot+1.35 M_\odot$ BNS, i.e.,
\begin{align}
A' = 2.5 \times 10^{-22} \times \frac{50 {\rm Mpc}}{d}\,.
\end{align}
While choosing $A'$ based on results from $1.35 M_\odot+1.35 M_\odot$
BNSs is not ideal, it should provide a reasonable approximation if
Eq.~\eqref{Mnsdistribution} is valid for merging BNSs, because it is
narrowly peaked around $1.33M_\odot$ and hence the majority of BNSs
are near equal mass binaries with total mass $\sim 2.7M_\odot$.
Nevertheless, such a prescription needs to be revised once we gain
more systematic (and accurate) understanding of the functional
dependence of $A'$ and $Q$ on the binary intrinsic parameters from
future numerical simulations of BNS mergers, complemented by actual
observations. Since in this section we are only interested in SNRs of
individual events, we will set the phase offset to zero.

\begin{figure}[tb]
\includegraphics[width=8.4cm]{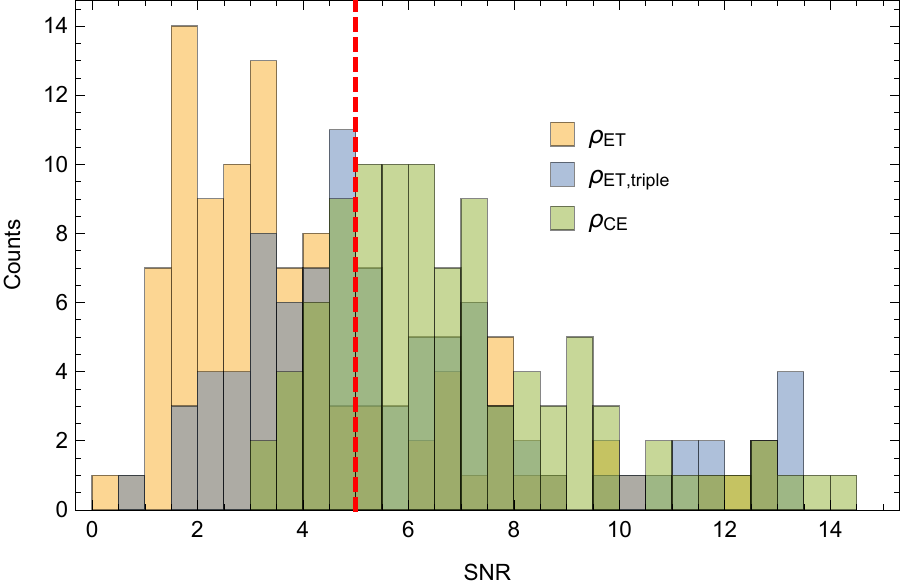}
\caption{Histogram for single event SNR for $100$ realizations in the
  MC simulation. Orange bins represent the SNR with respect
  to the sensitivity of the ET (single detector). Blue
  bins are associated with the triple-detector, triangle design of the
  ET \cite{regimbau2012mock}.  Green bins 
  represent
  the SNR with respect to the sensitivity of the CE
  (wide-band configuration). The detection threshold ($\rho=5$) is
  indicated by the red, dashed line.  The TM1 EOS and one year observation 
  is assumed, and the binary merger rate is taken
  to be $R_{BNS} = 1.54 {\rm Mpc}^{-3} {\rm Myr}^{-1}$.  }
\label{fig:plot1}
\end{figure}

\begin{figure}[tb]
\includegraphics[width=8.4cm]{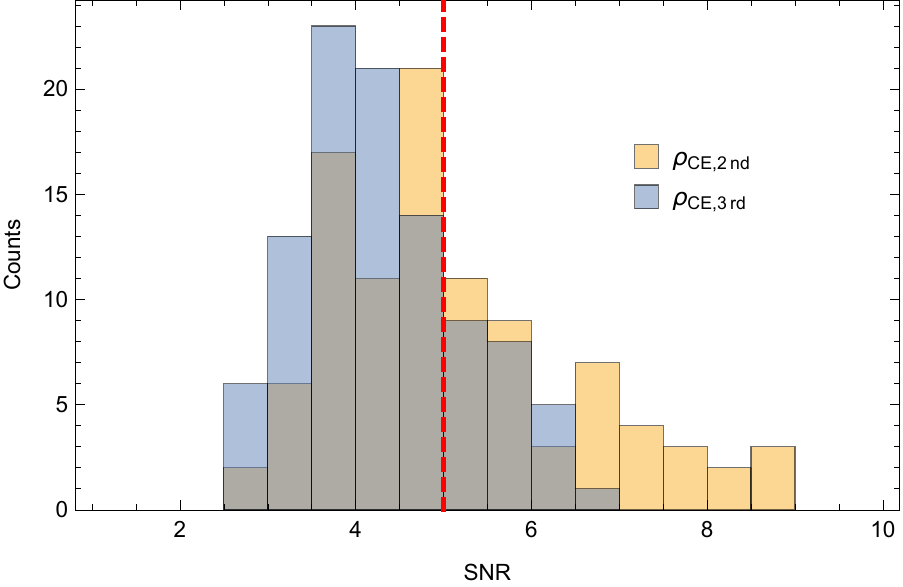}
\caption{Histograms of the second and third loudest events with the CE sensitivity, from
 the same MC runs as in Fig.~\ref{fig:plot1}.  Orange bins represent
  the SNR of the {\it second loudest} event, while
  blue bins represent the SNR of the {\it third loudest}
  event.}
\label{fig:plot2}
\end{figure}

We ran 100 MC realizations each covering one-year of observations to
calculate SNRs. In each realization, we reject binaries with total
mass exceeding the prompt-collapse threshold mass $M_{\rm thres}$
based on the results of~\cite{bauswein2013prompt}. We note that,
strictly speaking, a BNS with total mass just below $M_{\rm thres}$ in
general cannot survive for more than a few ms following merger, and
hence it cannot exhibit any loud post-merger oscillations. The
simulations of~\cite{Shibata:2006nm} suggest that the threshold mass
for rejecting binaries from our MC realizations should be $\sim
0.95\times M_{\rm thres}$. However, this effect has a small
contribution to our results because the total mass distribution for
BNSs derived from Eq.~\eqref{Mnsdistribution} is also Gaussian and is
given by Eq.~\eqref{Mnsdistribution} with $M_{0} = 2.66M_\odot$ and
$\sigma=0.1273M_\odot$. As a result, the fraction of binaries with mass above
$0.95\times M_{\rm thres}$ is only $1\%$ (for the TM1 EOS\footnote{For other EOSs we introduce later, this
fraction is practically the same for the LS220 EOS, even lower for the
DD2 and Shen EOSs, and a bit larger for the SFHo EOS (in the latter
case, however, we will see that the post-merger GWs are difficult to
detect in the first place).}).

For the single-event study we consider both the ET and CE
third-generation ground-based GW observatories.  The ET sensitivity is
obtained from \cite{amaro2009einstein} and the CE sensitivity from
\cite{abbott2017exploring}. For the CE, we choose the wide-band
configuration, because it has better sensitivity than the ``standard
configuration" above $1 {\rm kHz}$. For the ET configuration, we
consider both a single interferometer and a triangular arrangement
with three interferometers.

 In each MC realization there are about
40-70 events with $\rho >1$.  In Fig.~\ref{fig:plot1} we present the
SNR of the loudest event in each of the $100$ MC realizations with the
TM1 EOS. Our MC simulations show that for the ET (with single
interferometer) and CE sensitivities, there is a $25\%$ and $79\%$
chance respectively to have a single loud event passing the detection
threshold after acquiring data for a full year (for the
triple-detector ET case it is a $56 \%$ chance). In
Fig.~\ref{fig:plot2} we present the SNR of the second and third
loudest event in each of $100$ MC realizations for CE. The plot shows
that after a year of observations there is about a $43\%$ chance to
have a second and a $23\%$ chance to have a third event above
detection threshold.  The even lower chance of detecting secondary and
tertiary events above the detection threshold of 5 implies that there
is little room for stacking if one insists on using only signals above
this threshold in order to obtain a much stronger signal that will
reduce the statistical uncertainties in parameter estimation.
 
Given that the ET and CE are the most sensitive ground-based GW
detectors proposed so far, our results indicate that (for the
currently envisioned configurations) unless the true BNS merger rate
turns out to be substantially higher than the rate we adopt in our
study, or nature supplies a more ``favorable'' EOS as discussed in the next paragraphs,
over the next few decades the prospect to directly probe the
dominant peak of BNS merger remnant oscillations from {\em individual}
events does not appear very promising. Of course, this observation is a
consequence of adopting $\rho_{\rm thres}=5$. Since post-merger
oscillations will be an example of a {\em triggered} search, it is
conceivable that a lower threshold could be targeted.

All of the above results were obtained with the TM1 EOS, but we also
studied several other popular and realistic, finite temperature
nuclear EOSs. In particular, we considered the Steiner et al. EOS
SFHo~\cite{Steiner:2012rk}, the Lattimer Swesty
EOS~\cite{1991NuPhA.535..331L} with compressibility parameter $K=220$
MeV (LS220), the Hempel et al. EOS DD2~\cite{Hempel:2011mk}, and the
Shen et al. EOS~\cite{Shen:2011qu}. These EOSs were chosen because
they all have a maximum mass above
$2.0M_\odot$~\cite{Demorest2010,Antoniadis2013}, they cover a range of
stiffness, and because they take into consideration finite temperature
effects self-consistently. The parameters for performing the MC
simulations with these different EOSs are listed in
Table~\ref{table:para}. The strain amplitude $A'$ and the quality
factor $Q$ are chosen such that the peak value of the characteristic
strain and SNR of Eq.~\eqref{eq:w22} match the peak value of the
characteristic strain and SNR of the post-merger dominant $22$
component reported in the BNS merger simulations
of~\cite{Sekiguchi:2011mc,Stergioulas:2011gd,Palenzuela:2015dqa}. In
Appendix~\ref{appendix:fits} we show how well the Lorentzian profile
of Eq.~\eqref{eq:w22} approximates the post-merger spectra in the
vicinity of the dominant post-merger peak found in numerical
relativity simulations.

\begin{figure}[tb]
\includegraphics[width=8.4cm]{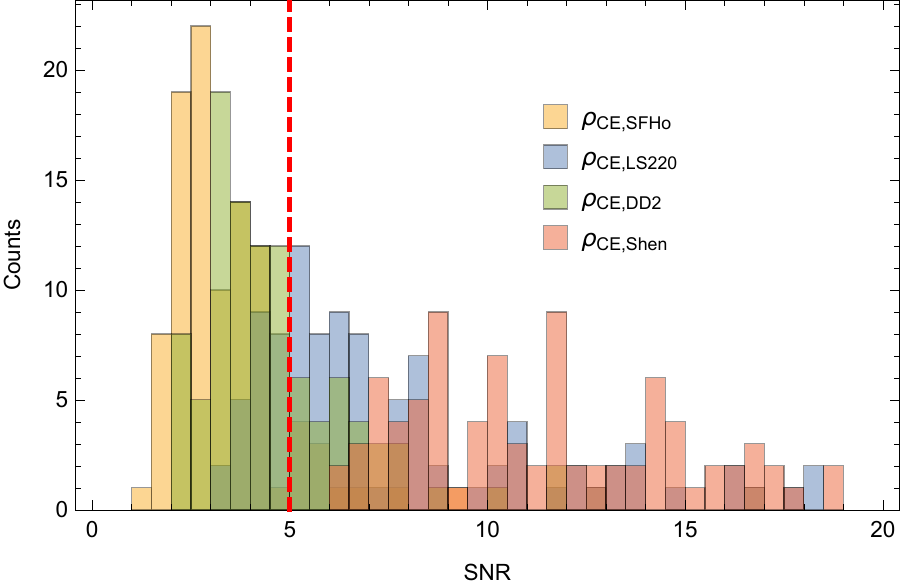}
\caption{The same setting as in Fig.~\ref{fig:plot1} but with different EOSs
  and with respect to the CE sensitivity alone.  Orange bins represent
  the SNR for the SFHo EOS, blue bins for the
  LS220 EOS, green bins for the DD2 EOS and red
  bins for the Shen EOS. }
\label{fig:plot3}
\end{figure}

Assuming $R_{BNS} = 1.54 {\rm Mpc}^{-3} {\rm Myr}^{-1}$, the results of
the MC realizations with different EOSs are presented in
Fig.~\ref{fig:plot3}, which shows that, among the EOSs that we study,
both SFHo and DD2 EOS have small detection rates ($\sim 13\%$ and $30 \%$ respectively, with $76\%$ for LS220 and $100\%$ for Shen)
for post-merger oscillations after a full year of observations with
CE. However, it should be stressed that these results should
be considered only as approximate, with the detailed numbers subject
to change with more accurate modeling of NS mergers in the coming
years.

Given the richness and importance of the physics encoded in the
post-merger signal, there is strong motivation to improve its
detectability by exploiting the information we can anticipate from the
current/planned generation of detectors, and informing designs for
future GW detectors to maximize their sensitivity to this phase of BNS
mergers. In this work we are focusing on the former approach, and in
next section show that stacking signals from multiple
detections can significantly enhance the sharpness of the post-merger
signal. We describe the details of the power and coherent mode
stacking methods we propose in the next section.

\begin{table}[ht]
\caption{Parameters for different EOS}
\centering
\begin{tabular}{c c c c c c}
\hline\hline
EOS & $R_{1.6 M_\odot}$ & $f_{\rm peak} (\rm kHz) \frac{M_\odot}{m_1+m_2}$ & $\frac{A' (50 {\rm Mpc})}{10^{-22}}$ & Q & $\frac{M_{\rm thres}}{M_\odot}$ \\
\hline
SFHo & 11.77 &1.21 & 2.7 & 25.7 & 2.95 \\
LS220 & 12.5 & 1.09 & 4.3 & 25.7 & 3.05 \\
DD2 & 13.26 & 0.98 &  2.8 & 12.7 & 3.35 \\
Shen & 14.42 & 0.84 & 5.0 & 23.3 & 3.45 \\
TM1 & 14.36 & 0.85 & 2.5 & 34.2 & 3.1 \\
\hline
\end{tabular}
\label{table:para}
\end{table}

\section{Multiple event detection}\label{sec3}

In this section we present the GLRT formalism for stacking multiple
events, and we assess the detectability of BNS post-merger oscillations
through such a stacking analysis using third-generation ground based
GW interferometers.

The GLRT formalism described in Section~\ref{sec21} justifies the
rationale of claiming detection from a single event with unknown
phase, within the Bayesian framework. When it is applied to
single-event detections, it should give consistent results with
previous studies~\cite{clark2016observing}.  However, as discussed in
Sec.~\ref{sec22}, these single events are not likely to allow a direct
detection of post-merger oscillations even when using the most
sensitive ground-based GW detectors proposed so far, unless event
rates are higher than expected or the EOS is Shen-like.
In the few cases where we manage to beat
the odds and have a loud event, the chance of also having a second
sufficiently loud event is even slimmer (Fig.~\ref{fig:plot2}).
Nevertheless, regardless of whether there is a loud event passing the
detection threshold, our MC studies indicate that there will likely be
several tens of events with modest SNR (e.g. $1\le \rho \le 5$), which
we exploit in this work to increase the chances of detection and
improve the accuracy of parameter estimation. Notice that although the
post-merger SNRs of these events are low, their inspiral SNRs will
be significantly higher and easily detectable.
This point will be further discussed in Sec.~\ref{sec31}.

\subsection{Hypothesis testing with power stacking}\label{sec:power}

In this section, we apply the Bayesian model selection approach discussed in \cite{meidam2014testing} for multiple events, which we refer to as 
power stacking.
With a group of events $y_i$ ($i=1,..,N$), the combined Bayes factor is \cite{meidam2014testing}
\begin{align}
B_{12} =\prod^N_{i=1} \frac{P( y_i | \mathcal{H}_1 )}{P( y_i | \mathcal{H}_2 )}\,,
\end{align}
where  the combined $\hat{T}$ variable is
\begin{align}\label{eqpowers}
2 \hat{T}_{\rm power}  \equiv & 2 \log \prod^N_{i=1} \left . \frac{P(y_i|\mathcal H_1)}{P(y_i|\mathcal H_2)} \right . \nonumber \\
=& \sum^N_{i=1} \left \{ \frac{(\langle c_i | y_i \rangle)^2}{ \langle c_i | c_i \rangle}+\frac{(\langle s_i | y_i \rangle)^2}{ \langle s_i | s_i \rangle}\right \} \nonumber \\
 =&\sum^N_{i=1} \langle c_i| c_i \rangle (A^2_{c,i}+A^2_{s,i})+\sum^N_{i=1} \frac{\langle c_i | n_i \rangle^2+\langle s_i | n_i \rangle^2}{\langle c_i | c_i \rangle } \nonumber \\
+& 2\sum^N_{i=1} \frac{A_{s,i} \langle s_i | n_i \rangle+A_{c,i} \langle c_i | n_i \rangle}{\langle c_i | c_i \rangle} \nonumber \\
=&2 \mathbf{s}_{Tp}+2 \mathbf{n}_{Tp}\,.
\end{align}

Note that in writing down such a combination of Bayes factors, we have implicitly 
assumed that all events follow the same hypothesis. This assertion relies on
the assumption that post-merger oscillations should exist if the mass of the 
remnants is below the (EOS-dependent) threshold.
This GLRT variable does not have the same type of noise distribution
as that analyzed in Sec.~\ref{sec21}.  Instead, its
distribution is obtained from a non-central $\chi^2_{2N}$ distribution
with an appropriate change of variable.  For sufficiently large N, due
to the central limit theorem, the distribution of $\mathbf{n}_{Tp}$ is a
Gaussian
with mean $N$ and variance (all $c_i, s_i$ are normalized such that
$\langle c_i | c_i \rangle = \langle s_i | s_i \rangle =1$)
\begin{align}
{\rm Var} [\mathbf{n}_{Tp}] = N+\sum^N_{i=1} (A^2_{c,i}+A^2_{s,i}) \equiv \sigma^2_{\mathbf{n}_{Tp}}\,.
\end{align}
The distribution of the noise ($2 \mathbf{n}_{Tp}$) 
associated with the null
hypothesis $A_{c,s} =0$ is just a $\chi^2_{2N}$ distribution, which
also asymptotes to a Gaussian distribution in the large $N$ limit with
mean $2 N$ and variance $4 N$.  Let us denote
\begin{align}
Q_\sigma (x) \equiv  \frac{1}{\sqrt{2\pi} \sigma} \int^\infty_x dy\,e^{-y^2/(2\sigma^2)} \,,
\end{align}
and 
\begin{align}
U_{2N} (x) \equiv \int^\infty_x dy\, P_{\chi^2_{2N}} (y)\,,
\end{align}
so that the requirement to reject the null hypothesis with significance level $P_f$, and 
the change of success (detection rate) is $P_d$, is
\begin{align}\label{eqpg}
\sum^N_{i=1} (A^2_{c,i}+A^2_{s,i}) \ge & U_{2N}^{-1} (P_f)-R_{2N}^{-1}(P_d)\nonumber \\
\approx &  U_{2N}^{-1} (P_f)-2N-2Q^{-1}_{\sigma_{\mathbf{n}_{Tp}}}(P_d)\,,
\end{align}
where $R_{2N}$ is the right-tail probability function for the
random variable $2 \mathbf{n}_{T_p}$. In practice, $P_d$
being around $0.99$ is already a decent detection rate.

\subsection{Hypothesis testing with coherent stacking}\label{sec31}

The coherent stacking approach developed in \cite{yang2017black} for black hole
ringdowns relies on extra information to align the phase between modes
in different events. For BNS mergers such accuracy in theoretical
modelling (even given the EOS) is unavailable at present. However,
with improvements in numerical simulations expected in the future, it
is possible that the inspiral part of BNS waveforms could be used to
predict the phase of post-merger modes (again, given the EOS). The
investigation in this section relies on the above assumption.

We develop a method of coherent stacking that relies on the existence
of a dominant post-merger peak frequency, universally related to the
component masses and the radius via Eq.~\eqref{eq:fp}. The component
masses and the time of merger can be determined from the inspiral part
of the waveform to high accuracy. For example, a Fisher analysis
suggests that the mass of each individual component in a $1.35
M_\odot+1.35M_\odot$ NS binary located at $300 {\rm Mpc}$ away from
Earth can be measured to a precision better than $\sim 0.1\%$,
assuming optimal sky location and orientation of the source using CE.
The relation in Eq.~\eqref{eq:fp} itself is not exact, but we expect
the theoretical understanding leading to it to improve over time with
more accurate numerical simulations and better constraints on the
EOS. Of course, one should note that even if Eq.~\eqref{eq:fp} were
exact for the set of candidate EOSs studied, there could still be a
systematic error if none of these EOSs are close enough to the true
finite temperature NS EOS. If this is the case, $f_\mathrm{peak}$ will
be erroneously predicted, degrading the efficiency of the coherent
stacking process.  We refer to Sec.~\ref{sec4a} for more discussions
on comparing different EOSs.

In our stacking approach we pick events with modest SNR ($\rho \ge 1$)
and assume that the phase $\phi^0$ can be determined by the inspiral
waveform within an uncertainty
\begin{align}
\label{eq:phase-uncertainty}
\sigma_{\phi^0} \approx \frac{C}{\rho_{\rm inspiral}}\,,
\end{align}
where $C$ is a constant to be determined by future simulations. For
the Monte-Carlo investigation in Sec.~\ref{sec32} we choose
$C=2\pi$.

Different events will in general have different remnant masses and
hence different 22 mode frequencies; therefore, we need to rescale the
data before stacking so that all $22$ modes have the same
frequency. Such a procedure has been described in detail in
\cite{yang2017black} to reprocess data before coherent stacking of BH
ringdown modes, and in \cite{clark2016observing} for constructing a
``universal" template bank.

With this at hand, let us now proceed with stacking. Suppose we have a
set of \emph{rescaled} data from $N$ different events with $\rho > 1$
\begin{align}
\tilde y_i(f) = \tilde g_i(f)+\tilde n_i(f)\,,
\end{align}
where $i=1,\ldots, N$ labels different events with $\tilde g_i \equiv
A_{c,i} \tilde h_{c,i}+A_{s,i} \tilde h_{s,i}$.  We further assume
that $\phi_i =\phi^0_i+\delta \phi_i$ is the estimator for the phase
of each event, where $\phi^0_i$ are the unknown, true underlying
phases while $\delta \phi_i$ is the measurement uncertainty of
$\phi_i$.  We then align the phases and coherently sum up the data
with different weights via ($0<w_i \leq 1$):
\begin{align}\label{eqsyny}
\tilde {\bf y} = &\sum_i w_i e^{i \phi_i} (\tilde g_i+\tilde n_i) \nonumber \\
= & \tilde {\bf g}_y+\tilde {\bf n}_y\,,
\end{align}
where $\tilde {\bf g}_y$ ($\tilde {\bf n}_y$) is the signal (noise) part of the stacked data $\tilde {\bf y}$.

The stacked data can now be used to construct the log Bayes factor for
the hypothesis that a signal is present versus one where no signal is
present.  Based on the discussion in Sec.~\ref{sec21}, we then need to
evaluate the quantity $\langle {\bf y} | \mathcal{I} | {\bf y}
\rangle$, with $\mathcal{I} \equiv | c\rangle \langle c |+|s \rangle
\langle s |$ and the brackets $\langle | \rangle$ here are defined
with respect to the spectrum of $n_y$. The quantities $c$ and $s$ are
again defined as in Sec.~\ref{sec21} but with individually rescaled
frequencies.  It is also straightforward to verify that $\mathcal{I}
|g_i \rangle =| g_i \rangle$, because we assume that the frequency
uncertainty with a {\it known} EOS is negligible. Using this property,
$\hat{T}_{\rm coherent}$ is given by
\begin{align}
2 \hat{T}_{\rm coherent} &\equiv \langle {\bf y} |\mathcal{I} | {\bf y} \rangle  \nonumber \\
&=  \sum_i w^2_i \langle g_i    | g_i \rangle+\sum_{i \neq j} w_i w_j \langle g_i  e^{i \phi_i} | g_j e^{-i \phi_j}  \rangle\,\nonumber \\
& +\sum_{ij} w_i w_j \langle n_i e^{i \phi_i}   | \mathcal{I} |n_j e^{-i \phi_j}  \rangle  \nonumber\\
&+ \sum_{ij} w_i w_j [\langle g_i e^{i \phi_i}  | n_j  e^{-i \phi_j} \rangle +\langle n_i  e^{i \phi_i}  | g_j  e^{-i \phi_j} \rangle ]\,\nonumber \\
& = 2 {\bf s}_{Ty}+2 {\bf n}_{Ty}\,,
\end{align}
where $2{\bf s}_{Ty}$ is used to designate the term appearing on the
first line and $2{\bf n}_{Ty}$ all remaining terms. We refer to ${\bf
  s}_{Ty}$ (${\bf n}_{Ty}$) as the signal (noise) part of
$\hat{T}_{\rm coherent}$.

The signal part of the stacked data can now be used to determine a
detection criterion.  We begin by evaluating ${\bf s}_{Ty}$ with an
ensemble average over the phase uncertainties (using $\langle e^X
\rangle =e^{-\langle X^2 \rangle /2} $ for any Gaussian random
variable $X$ with zero mean)
\begin{align}\label{eqts}
2 \langle {\bf s}_{Ty} \rangle =&\sum_i w^2_i \langle g_i | g_i \rangle \nonumber \\
 &+\sum_{i \neq j} w_i w_j \langle g_i  e^{i  \phi^0_i} | g_j  e^{-i \phi^0_j} \rangle  e^{-\frac{\sigma^2_{\phi_i}}{2}-\frac{\sigma^2_{\phi_j}}{2}}\,,
\end{align}
which corresponds to the stacked SNR squared, and
where $\sigma^2_{\phi_i} =\langle \delta \phi_i^2 \rangle$. Based on
the discussion in Sec.~\ref{sec21}, the SNR of $y$ has to be larger
than $5$ to pass the detection threshold. Then, the detection criteria
for the stacked signal is just
\begin{align}\label{eqcsreq}
\sqrt{2 \langle {\bf s}_{Ty} }\rangle \ge 5\,.
\end{align}
The weight coefficients $w_i$ are chosen such that $\langle {\bf
  s}_{Ty} \rangle$ is maximized, and in this work it is achieved 
using the downhill simplex optimization method
~\cite{Nelder1965, Press:2007nr}. Similar to the single event case, we 
present the variance of ${\bf n}_{Ty}$ in Appendix~\ref{appendixA}, together
with the signal-to-noise level of $\hat{T}_{\rm coherent}$.
  
The performance of stacking is discussed in Sec.~\ref{sec32}, but let
us make an immediate observation. If there are $N$ events under
stacking and all of them have comparable SNR, this coherent stacking
method would produce an ${\cal{O}}(N^{1/2})$ boost in ${\bf
  s}_{Ty}$\footnote{Notice that $S_n$ in the definition of the inner
  product scales linearly with $N$.}. In reality, there is always a
small group of events with high SNR, while the remaining events have
low SNR. Thus, in practice the improvement factor over the event with
{\em best} SNR can never achieve $N^{1/2}$-type scaling. The same
observation was made when coherently stacking ringdown modes from BH
coalescences \cite{yang2017black}.

\subsection{MC study}
\label{sec32}

In this section, we show how stacking enhances the chance of detecting
BNS post-merger signals by using the results of our MC simulations. We
first compare the results for power stacking against single event
detection. We next compare coherent stacking against power stacking
and show that the former works more efficiently than the latter.

We note that it is difficult to define a SNR for a combined set of
events, because the statistical distributions of $\hat{T}$ for the true and
null hypotheses ($\mathcal{H}_{1,2}$) are different from those of a
single event (see Eq.~\eqref{eqso}). As a result, we define
a new quantity $\alpha$, which is the universal scale factor that the
SNR of all events should be divided by (or detector noise should be
multiplied by): $A_{i} \rightarrow A_i/\alpha$, in order to exactly
satisfy the detection bound in Eq.~\eqref{eqpg} or
Eq.~\eqref{eqcsreq}. The larger this detection-threshold-matching
factor $\alpha$ is, the more efficiently an analysis method
performs. We shall apply this $\alpha$ to characterize the performance
of stacking in this Section.

\subsubsection{Power stacking versus single event detection}
\label{sec:stacking-vs-single}

\begin{figure}[tb]
\includegraphics[width=8.4cm]{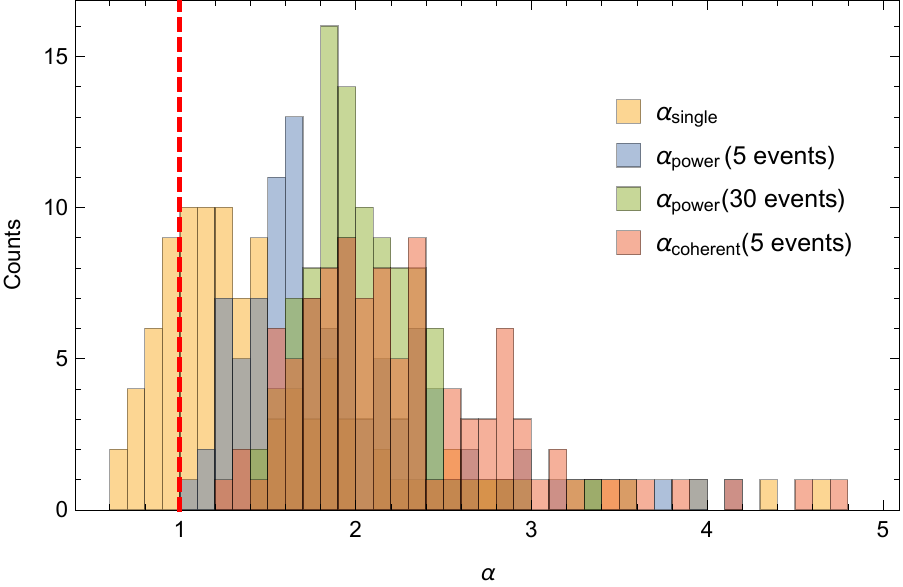}
\caption{
  Histogram for power-stacked, coherently stacked and single-event
  $\alpha$ (a proxy for SNR, with $\alpha=1$ being the detection threshold),
  in each realization for the
  TM1 EOS. Orange bins represent the top event in each MC realization without stacking, with
  $79/100$ passing the detection threshold. Blue bins
  represent power stacking using the loudest $5$ events, and
  demonstrate that all realizations pass the detection threshold.
  Green bins represent power stacking using the loudest
  $30$ events, showing an improvement compared
  to power stacking $5$ events.
  Pink bins represent coherently stacking the top $5$ events
  with $C=2\pi$ in Eq.~\eqref{eq:phase-uncertainty}; all cases
  pass the detection threshold, and the skew of the distribution 
  toward larger values of $\alpha$ 
  indicates that
  coherently stacking the top 5 events is more efficient than
  power stacking the top 30 events.}
\label{fig:alpha}
\end{figure}

In each MC realization performed in Sec.~\ref{sec22}, we pick the top
$N$ events to construct the Bayes factor in Eq.~\eqref{eqpowers}.  In
Fig.~\ref{fig:alpha}, for illustration purposes we choose $N$ to be $5$
and $30$, although $N$ can be any
positive integer less than or equal to the total number of events
detected in general.

 \begin{figure}[tb]
\includegraphics[width=8.4cm]{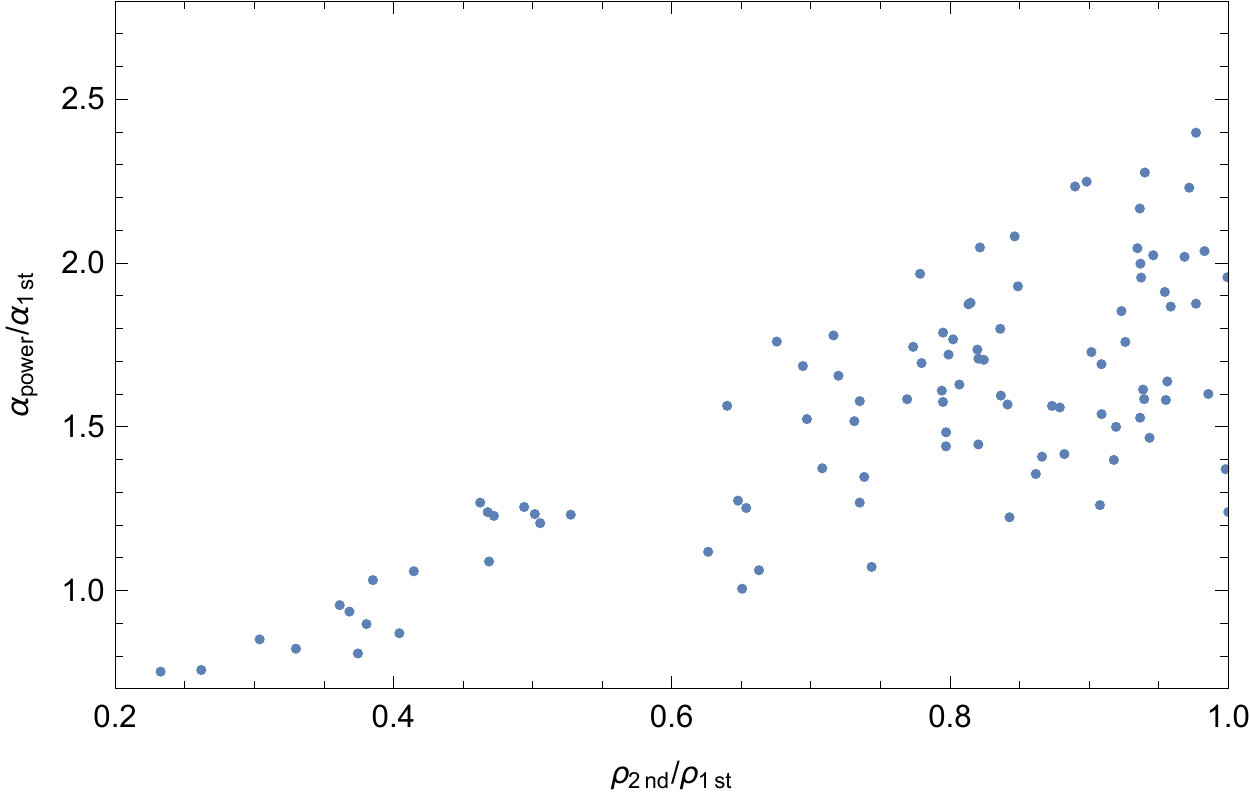}
\caption{Vertical axis: the improvement factor of
  effective SNR for the power stacked signal ($N=30$) over the best single
  event in each MC realization with the TM1 EOS. Horizontal axis: the
  ratio of $\rho$ between the second best event and the best event in
  each MC realization.  }
\label{fig:plotscatter}
\end{figure}
 
In Fig.~\ref{fig:alpha}, $\alpha$ for any single event is equal to its
${\rm SNR}/5$. Before applying power stacking with the TM1
EOS, there is roughly a $79\%$ chance to detect a post-merger $22$
mode with CE operating for a year.  Power stacking leads to a
decrease in the false alarm rate. For $N=5$, this decrease is enough
to allow the stacked signals to pass the detection threshold in all MC realizations. For $N=30$, all stacked events
are able to pass the detection threshold, and the improvement factor in $\alpha$ is roughly a factor of $1.5$. In principle, events with
low SNR could contribute more noise fluctuations than signal
improvement in $\hat{T}_{\rm power}$, which means that adding more
events does not necessarily lead to better statistics for detection.
This is shown in Fig.~\ref{fig:plotscatter} because some of the MC
realizations have $\alpha_{\rm 1st} > \alpha_{\rm
  power}$. Here $\alpha_{\rm 1st}$ stands for the $\alpha$ factor
  of the top event in each MC realization.

Intuitively we can interpret $5 \times \alpha_{\rm power}$ as the
``effective SNR" of the power stacked signals, as $5 \times
\alpha_{\rm single}$ is the SNR of a single event. In this sense,
$\alpha_{\rm power}/\alpha_{\rm 1st}$ just characterizes the
improvement in effective SNR by power stacking. As shown in
Fig.~\ref{fig:plotscatter}, this effective SNR improvement ranges
between $0.7$ and $2.5$, with median value at $1.57$. When the SNR of
the best event is much higher than the rest, so that $\rho_{\rm
  2nd}/\rho_{\rm 1st}$ is small, the effective SNR improvement tends
to be smaller. Therefore the power stacking approach works better for
events with more uniformly distributed SNRs.

We conclude this subsection with a short discussion of how our results
would change if we had chosen a different EOS.  Assuming that the
enhancement in SNR (a factor of $\sim 1.5$) due to power stacking
relative to a single event does not depend strongly on the EOS, one can
roughly estimate the distribution of $\rho$ after stacking for various
EOSs by shifting the histograms in Fig.~\ref{fig:plot3} to larger SNR
by a factor of $\sim 1.5$.  Doing so, one finds that it is very likely
that the stacked signal can be detected for all EOS we consider here
except for the case of the SFHo EOS. This clearly shows that the
detectability of the post-merger signal is sensitive to the underlying
EOS.

\subsubsection{Coherent stacking versus power stacking}

We now compare the SNR improvement between coherent
and power stacking. In
Fig.~\ref{fig:alpha} we show $\alpha$ for a coherently stacked
signal using the top $5$ events in each of the $100$ MC realizations discussed
earlier, and $C$ in (\ref{eq:phase-uncertainty}) is assumed to be $2\pi$.  Because the inspiral SNRs of
these events are much greater than the post-merger SNR, the effect due
to phase and frequency uncertainties of modes is negligible. The
weight $w_i$ is basically proportional to post-merger ${\rm SNR}_i$,
and the SNR of the coherently stacked signal to close to $\sqrt{\sum_i
  {\rm SNR}^2_i}$. Figure~\ref{fig:alpha} demonstrates that the
$\alpha$ distribution for coherent stacking of \emph{five events}
is skewed toward larger values than power stacking of \emph{thirty
  events}. Thus, coherent mode stacking outperforms power stacking in
this setting. 

Another way to compare the two stacking approaches is to consider a
simple scenario in which all $N$ events have identical SNR, and ask
how many events are needed to satisfy the detection threshold for
each stacking method. For power stacking, this gives the following equation for the threshold
number $N$ in terms of the individual event SNR $\rho$ (for
simplicity, we ignore the fact that $N$ has to be integer)
\begin{align}\label{eqpowerna}
\frac{\sqrt{N}}{2} \rho^2 &= \frac{1}{2 \sqrt{N}} [U_{2N}^{-1} (P_f) -2N]-\sqrt{1+\rho^2} Q^{-1}_1(P_d) \nonumber \\
& \approx Q_1^{-1} (P_f) -\sqrt{1+\rho^2} Q^{-1}_1(P_d)\,.
\end{align}
 One immediate observation is
that unlike the coherent stacking case discussed in Sec.~\ref{sec31},
the $N - \rho$ relation is not a single power-law. For example, if $P_d=0.5$, 
the second term in Eq.~\eqref{eqpowerna} vanishes and we
can see that the threshold SNR satisfies $\rho \propto N^{-1/4}$. On
the other hand, if the second term dominates over the first term in
Eq.~\eqref{eqpowerna} and $\rho \gg 1$, the threshold SNR satisfies
$\rho \propto N^{-1/2}$.

\begin{figure}[tb]
\includegraphics[width=8.4cm]{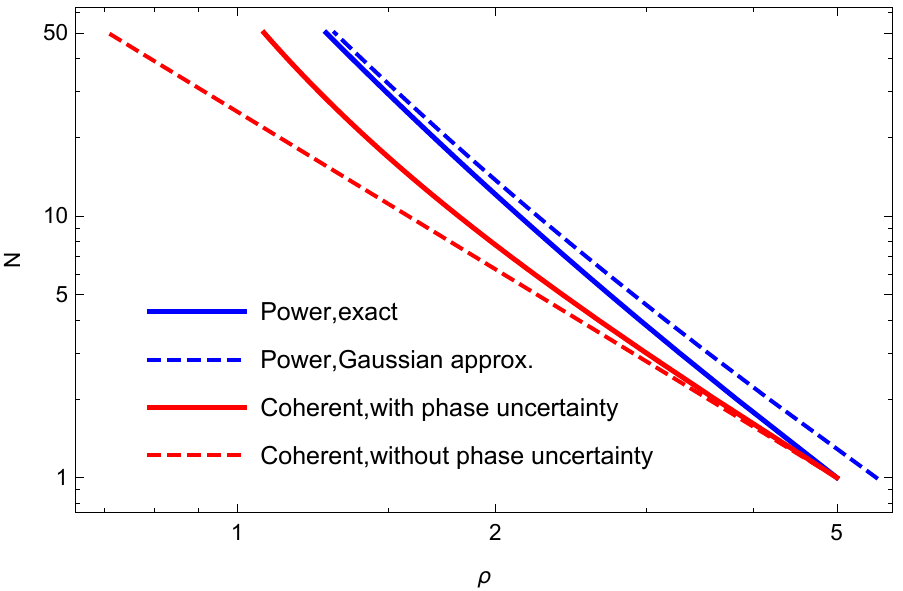}
\caption{ Number of identical events needed to satisfy the detection
  threshold, as a function of single event SNR $\rho$, with
  the false alarm rate $P_f=0.01$ and the detection probability
  $P_d=0.982$. The blue solid (dashed) line represents the
  requirement for power stacking with (without) the Gaussian
  approximation. The red solid (dashed) line represents the requirement
  for coherent mode stacking with (without) phase alignment
  uncertainty considered. Note that coherent stacking is always more
  efficient than power stacking (fewer events are needed to cross
  above the detection threshold).}
\label{fig:na}
\end{figure}

To compare the performance between coherent mode stacking and power
stacking, assuming all events have the same single SNR $\rho$, we
compute the number of identical-SNR events $N$ needed to satisfy the
equality in Eq.~\eqref{eqcsreq} for coherent stacking, and the
equality in Eq.~\eqref{eqpg} or Eq.~\eqref{eqpowerna} for power
stacking. In the coherent stacking case, this can be computed exactly:
\begin{align}
\sqrt{N} \rho = 5\,,
\end{align}
without considering phase uncertainty and
\begin{align}
  \label{coh_stack_Nrho}
\rho^2 [ 1+(N-1)e^{-\sigma^2_{\phi^0}}] = 25\,,
\end{align}
where for illustration purpose we also include a case with phase
uncertainty $\delta \phi^0 \approx 1/\rho$ \footnote{For a general
  event we expect $\rho$ to be proportional to $\rho_{\rm inspiral}$,
  so that $\delta \phi^0 \propto 1/\rho$ according to
  Eq.~\eqref{eq:phase-uncertainty}. We arbitrarily picked a coefficient $1$ to
  illustrate the effect of phase uncertainty.}.  In the power stacking
case, one must carry out the calculation numerically for a given
$P_{f}$ and $P_{d}$, as shown in Fig.~\ref{fig:na}. In this idealized
scenario, coherent stacking always outperforms power stacking as it
requires fewer events to pass the detection threshold for the same
$P_f =0.01$ and $P_d=0.982$. The Gaussian distribution approximation
(the second line in Eq.~\eqref{eqpowerna} corresponds to the blue
dashed line, which underestimates the performance of power stacking
(blue solid line) for small $N$, but agrees better with the exact
expression in the first line of Eq.~\eqref{eqpowerna} for larger $N$,
as expected. We also find that the phase uncertainty in the coherent
stacking case becomes more important in the low-$\rho$ regime -- the
red solid line departs more from the red dashed line -- as expected
from Eq.~\eqref{coh_stack_Nrho}.

\section{Model selection and Parameter Estimation}\label{sec4}

We now discuss how well one can distinguish two different EOS models, mainly focusing on power stacking.
We also show how accurately one can measure the peak frequency (and in turn 
the NS radius) with the stacked events.

\subsection{Model  selection for EOSs}\label{sec4a}

In order to compare the likelihood of different EOSs based on the
measured data, and in particular based on the stacked signal, we
perform a Bayesian model selection method to evaluate the relative
performance between different models. In general, for a given data set 
$y$ and two possible models $\mathcal{H}_1, \,\mathcal{H}_2$,
one can evaluate the Bayes factor given in Eq.~\eqref{eq:BF}\footnote{For coherently stacking multiple events, we generically obtain different stacked
signals for different EOSs, and thus different data sets that one has to compare to
perform model selection. This introduces a subtlety that we discuss in
Appendix~\ref{appendixB}, but for simplicity we ignore it in the following analysis.}.

Given that we can perform the analysis we present in Sec.~\ref{sec31}
for multiple EOSs, we can determine the EOS which gives the best SNR, which
here we call model 1, and then perform a model selection test for
other EOSs by using the data set $y$ corresponding to
model 1. Then within the GLRT framework, we evaluate the following
Bayes factor:
\begin{align}\label{eqodd3}
\mathcal B_{1|2} \equiv \frac{P(y | \mathcal{H}_1)}{P(y | \mathcal{H}_2)}\,.
\end{align}

We shall denote the two basis functions in model 1 as $c^{(1)}$ and
$s^{(1)}$ and the basis functions in model 2 as $c^{(2)}$ and
$s^{(2)}$. According to Eq.~\eqref{eqprob}, we then have
\begin{align}
\hat{\mathcal T}_{1 | 2} &= \log \mathcal B_{1|2} \nonumber \\
&=-\frac{|| y- \hat{A}_{1c} c^{(1)}-\hat{A}_{1s} s^{(1)}||^2}{2} \nonumber \\
&+\frac{|| y- \hat{A}_{2c} c^{(2)}-\hat{A}_{2s} s^{(2)}||^2}{2}\,.
\end{align}
Inserting the expressions for the maximum likelihood estimators
(c.f. Eq.~\eqref{eqmaxl}) in the above equation, we obtain 
\begin{align}
\hat{\mathcal T}_{1 | 2}  &=\frac{(\langle c^{(1)} | y \rangle)^2}{ 2\langle c^{(1)} | c^{(1)} \rangle}+\frac{(\langle s^{(1)} | y \rangle)^2}{ 2 \langle s^{(1)} | s^{(1)} \rangle}\, \nonumber \\
&-\frac{(\langle c^{(2)} | y \rangle)^2}{ 2\langle c^{(2)} | c^{(2)} \rangle}-\frac{(\langle s^{(2)} | y \rangle)^2}{ 2 \langle s^{(2)} | s^{(2)} \rangle}\, \nonumber \\
& =\Delta {\bf s}_{Ty} +\Delta {\bf n}_{Ty}\,,
\end{align}
where $\Delta {\bf s}_{Ty}$ and $\Delta {\bf n}_{Ty}$ are the signal and noise part of $\hat{\mathcal T}_{1 | 2}$ respectively.
For multiple events under the power stacking framework, we simply multiply all the posterior distributions, and the total Bayes factor is 
\begin{align}
\mathcal{T}_{1 | 2, {\rm power}} &= \log \prod^N_{i=1} \mathcal{B}_{1|2,i} \nonumber \\
&=-\sum^N_{i=1} \frac{|| y_i- \hat{A}_{1c,i} c_i^{(1)}-\hat{A}_{1s,i} s_i^{(1)}||^2}{2} \nonumber \\
&+\sum^N_{i=1} \frac{|| y_i- \hat{A}_{2c,i} c_i^{(2)}-\hat{A}_{2s,i} s_i^{(2)}||^2}{2}\nonumber \\
& =\Delta {\bf s}_{Tp} +\Delta {\bf n}_{Tp}\,.
\end{align}
In this case, the expectation of $T_{1 | 2}$ is 
\begin{align}
\label{eq:logB-power}
\langle \Delta {\bf s}_{Tp} \rangle  &= \sum^N_{i=1} \langle g_i | \mathcal{I}_{1,i} -\mathcal{I}_{2,i} | g_i \rangle \nonumber \\
&:= \langle \log \mathcal B_{1|2} \rangle_{,\mathrm{power}}\,.
\end{align}

On the other hand, we could coherently stack data from different events, if the assumption made in
Sec.~\ref{sec31} is met. By assuming that model 1 is the true EOS, we then find
\begin{align}
\label{eq:logB-coherent}
2 \langle \Delta {\bf s}_{Ty} \rangle &=\sum_i w^2_i (\langle g_i | g_i \rangle -\langle g_i |\mathcal{I}_2 | g_i \rangle )\, \nonumber \\
&+\sum_{i \neq j} w_i w_j \langle g_i  e^{i  \phi^0_i} | \mathcal{I}_1-\mathcal{I}_2|g_j  e^{-i \phi^0_j} \rangle e^{-\sigma^2_{\phi_i}/2-\sigma^2_{\phi_j}/2}\, \nonumber \\
& := \langle \log \mathcal B_{1|2} \rangle_{,\mathrm{coherent}}\,,
\end{align}
where we use $\langle \Delta {\bf s}_{Ty} \rangle$ as the expectation of $\log \mathcal{B}_{1 | 2}$ 
for coherent stacking. Notice that $2 \langle \Delta {\bf s}_{Ty} \rangle $ 
above reduces to $2\langle {\bf s}_{Ty} \rangle$ in Eq.~\eqref{eqts} 
when $\mathcal{I}_2=0$ (i.e. when $c^{(2)} = 0 = s^{(2)}$).

One can use the {\it Jeffreys scale of interpretation of Bayes Factor} \cite{jeffries1961theory} to determine how significant a Bayes factor is. If $ \mathcal{B}_{1 | 2}$ is between $[1,3]$, the statistical significance is barely worth mentioning; if $3 <  \mathcal{B}_{1 | 2} <10$ the evidence is strong; if $10 < \mathcal{B}_{1 | 2} <100$ the evidence is very strong and beyond $100$ it is decisive.

While events under detection threshold can be used to accumulate
statistics via stacking in Eq.~\eqref{eqso}, the interpretation of the
results of model selection (and also parameter estimation to be
discussed in Sec.~\ref{sec4b}) should be used with caution.  This is
because if the combined statistics of a set of events does not pass
the detection threshold, the existence of a $22$ mode in any of these
events is not confirmed. 
For simplicity we only
present the distribution of $\langle \log \mathcal B_{1|2}
\rangle_{,\mathrm{power}}$ (using the $30$ loudest events) versus the distribution of $\langle \log
\mathcal B_{1|2} \rangle$ for single events; repeating
the analysis with coherent stacking will introduce
the additional complication of dealing with the effect
of phase uncertainty, which we leave to future studies.

\begin{figure}[tb]
\includegraphics[width=8.4cm]{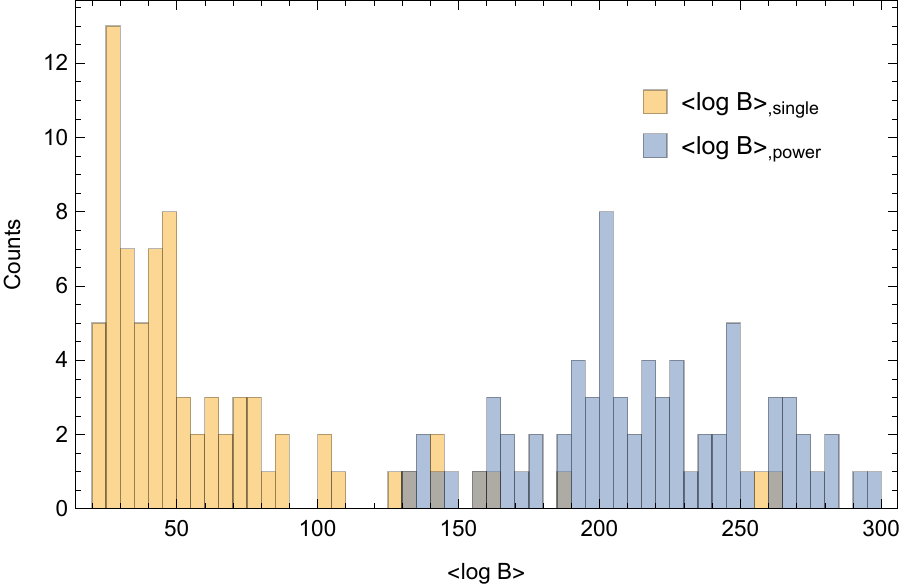}
\caption{Average (log of) Bayes factor for different MC realizations,
  testing EOS TM1 against DD2. Blue bins represent the stacked
  signals with the loudest 30 events according to Eq.~\eqref{eq:logB-power}. There are $79$ realizations
  where their best events individually pass the detection threshold, and their
  corresponding Bayes factors are shown in orange bins. }
\label{fig:modeltest}
\end{figure}

As an application, we assume TM1 to be the underlying EOS (model 1),
and test it against the EOS DD2 (model 2). As discussed in
Sec.~\ref{sec3}, there are $79$ out of $100$ MC realizations with at least one event passing the detection threshold, and all $100$ MC realizations that pass
the detection threshold when we apply power stacking for the top $30$ events. For each MC realization that can claim a detection, we compute the corresponding Bayes factor, which is shown
in blue bins in the histogram in Fig.~\ref{fig:modeltest}.  
Notice that 
these average Bayes factors can only be used to rank the models in a
semi-quantitative manner, as it is non-trivial to convert them to
probability measures. A full analysis would require one to generate  
statistical distributions of these Bayes factors for each underlying EOS. 
In other words, Fig.~\ref{fig:modeltest} should be interpreted as the scattering of the average Bayes factor due to the astrophysical distribution of sources. Based on the results of comparing a single pair of EOSs, we conjecture that as long as a
single event has passed the detection threshold, it can be used to
distinguish between different EOSs with ``very strong evidence''. However, single
events are less likely to be detected than the stacked events, and
thus, the latter have more chance of distinguishing different EOSs
than the former. 

\subsection{Parameter estimation for the peak frequency}\label{sec4b}

Given a  signal, we can also study the degree to which we can
estimate its peak frequency, which we do here via a Fisher analysis.
This can also serve as an alternative approach to distinguish between
different EOSs, as they generally predict different peak
frequencies. As a simple example, we assume TM1 as the best-fit EOS
and construct the stacked signal accordingly. Next we promote the
waveform (Eq.~\eqref{eq:w22}) parameter vector to four-dimensions:
\begin{equation}
\lambda^i = \left(A, \phi^0, f_{\rm peak}, Q\right)
\end{equation}
with $A = A' A_r$,  and maximize
Eq.~\eqref{eqprob} to obtain maximum likelihood estimators of these
parameters. 

In the Fisher approximation, the uncertainty in $\lambda^{i}$ can be
evaluated through the (Fisher) information matrix
\begin{align}
\Gamma_{ij} = \langle \partial_i h | \partial_j h\rangle\,,
\end{align}
where $\partial_i \equiv \partial/\partial \lambda^i$ and the
inner product is defined with respect to the spectrum of $\tilde {\bf
  n}_y$ in Eq.~\eqref{eqsyny}. The measurement uncertainty of $f_{\rm peak}$ is simply
\begin{align}
\delta f_{\rm peak} \geq \sqrt{\left(\Gamma^{-1} \right)_{f_{\rm peak} f_{\rm peak}}}\,,
\end{align}
where the right hand side corresponds to the square root of 
the $(f_{\rm peak},f_{\rm peak})$ element of the
variance-covariance matrix.  The inequality in the above equation
comes about because of the Cramer-Rao bound, which guarantees a
best-case measurement for a set of parameters in the high SNR
limit~\cite{Vallisneri:2007ev}.  We will use a Fisher analysis here
only as a rough estimate of the accuracy to which parameters can be
measured; a more complete analysis would construct the posterior
probability distribution for each parameter through a detailed mapping
of the likelihood surface, but this is beyond the scope of this paper.

Using these arguments and the approximations
in~\cite{clark2016observing}, we assume that the off-diagonal terms of the
$\Gamma$ matrix are small, 
so that
\begin{align}
\label{eq:delta-f}
\delta f_{\rm peak} \approx \left(\Gamma_{f_{\rm peak}f_{\rm peak}}\right)^{-1/2} &=\langle
\partial_{f_{\rm peak}} h | \partial_{f_{\rm peak}} h \rangle^{-1/2} \nonumber \\
& \approx 0.7\frac{f_{\rm peak}}{Q \,\rho}\,,
\end{align}
where in the last approximate equality we used the fact that the
Fourier transform of Eq.~\eqref{eq:w22} satisfies $\partial_{f_{\rm
    peak}} \tilde{h} \sim Q\tilde{h}/f_{\rm peak}$.  The factor of
$0.7$ comes from a numerical fit to our set of data using the TM1 EOS,
which is also expected by computing $\delta f_{\rm peak}/f_{\rm peak}$
for a universal Lorentzian-Type waveform. This shows that for $Q=34$
(corresponding to the TM1 EOS) with a signal of $\rho \sim 6.5$,
$f_\mathrm{peak}$ can be measured to $\sim 0.3$\% accuracy at best.

\begin{figure}[tb]
\includegraphics[width=8.4cm]{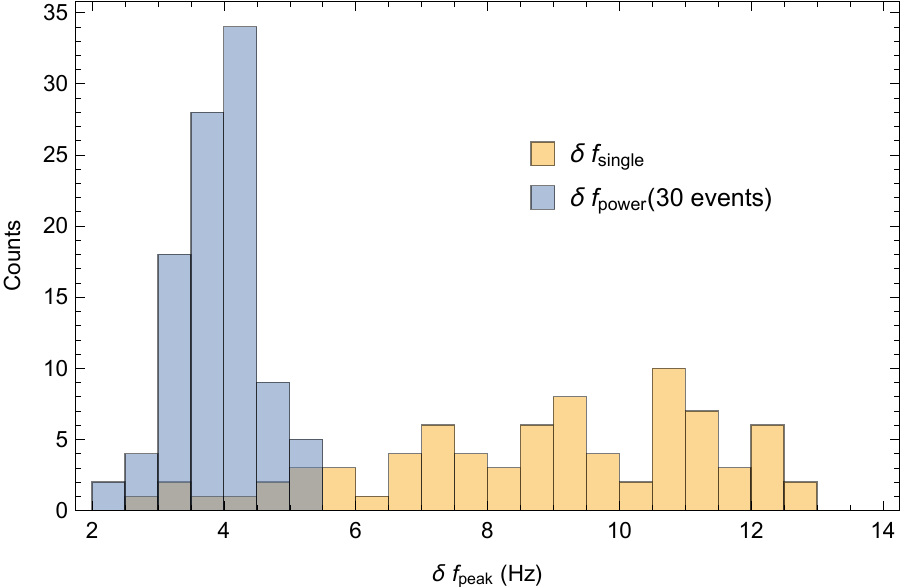}
\caption{ Histogram of $\delta f_{\rm peak}$ for different MC
  realizations, assuming the TM1 EOS and a $1.35+1.35 M_\odot$ BNS merger remnant with
  corresponding $f_\mathrm{peak} = 2.3$ kHz. The blue bins
  correspond to the power-stacked signal using the loudest 30 events from
  each of the $100$ realizations that pass the detection threshold, while the orange bins correspond to the
  79 individual events that pass the detection threshold.  A $\delta
  f_{\rm peak} \sim 8$ Hz measurement roughly corresponds to a $\sim
  26$ m statistical error in the determination of the radius of a NS
  with mass $1.6M_\odot$. However,
  systematic errors in the universal $f_\mathrm{peak} -
    R_{1.6M_\odot}$ relation used in the mapping could be larger than
  $\sim 100$ m, so these errors would dominate over the statistical 
  measurement error. }
\label{fig:df}
\end{figure}

We now look at the effect of the off-diagonal terms in the Fisher matrix.
If we include $\phi^0$ in addition
to $f_{\rm peak}$ into the Fisher analysis, as they both
enter the argument of the phase factor in the waveform, we find that the statistical
frequency uncertainty increases by a factor of $1.4$. Consequently,
the statistical radius uncertainty also increases by
$1.4$. Therefore, our Fisher analysis for TM1 including $\phi^0$ predicts $\delta f_{\rm
  peak}$ $\sim 8$ Hz.
For all the realizations with the best
signal passing the detection threshold, we evaluate the uncertainties
in $\delta f_{\rm peak}$ numerically, shown in
Fig.~\ref{fig:df}. 
  
Let us now map the statistical error in the peak frequency $\delta
f_\mathrm{peak}$ to that of the radius of a NS with mass $1.6 M_\odot$,
$\delta R_{1.6 M_\odot}$. From Eq.~\eqref{eq:fp}, one finds the
following relation:
\begin{equation}
\label{eq:deltaR}
\frac{\delta R_{1.6 M_\odot}}{1\mathrm{km}} = \frac{\delta f_\mathrm{peak}}{1\mathrm{kHz}} \frac{M_\odot}{m_1+m_2} \left [2 a_2 \left(\frac{R_{1.6 M_\odot}}{1 {\rm km}} \right) +a_1 \right ]^{-1}\,,
\end{equation}
where we have neglected the error in the estimation of the component
masses, as this is negligible for third generation detectors. Thus, a
$\sim 8$ Hz statistical uncertainty in the peak frequency roughly
corresponds to a $\sim 30$ m ($0.3\%$, ``TM1" EOS) uncertainty in
radius (for a NS with mass $1.6 M_\odot$). The total error
is the root of the sum of the squares of the statistical 
and all systematic errors. One source for the latter, as
discussed in \cite{clark2016observing}, comes from the $f_{\rm peak} -
R_{1.6 M_\odot}$ relation, and currently is above
$100$ m for $R_{1.6 M_\odot}$. Therefore, for now, the error budget is
dominated by systematic error and not statistical when considering
third generation detectors.  Of course, as discussed earlier, we
expect this systematic uncertainty to be considerably lowered by the time third
generation detectors come online, as more accurate understanding and
modeling of BNS merger remnants is developed.
  
Notice that the uncertainty in frequency, with mean at about $8$
Hz in Fig.~\ref{fig:df} is significantly smaller than the value of order $50 {\rm Hz}$ presented in
\cite{clark2016observing}.  
In fact, we notice that $\delta f_{\rm peak}/f_{\rm peak}$ in
\cite{clark2016observing} does not follow the $0.7 /(Q \rho)$ relation
derived here. In \cite{clark2016observing} for the TM1 EOS, $\delta
f_{\rm peak}/f_{\rm peak} \sim 50/2300$ with the post-merger SNR being
$5$. Based on Fig. 11 of \cite{clark2016observing}, the SNR of the
dominant $22$ component is roughly half of the post-merger total SNR,
i.e., $\rho \sim 2.3$. This converts to $\delta f_{\rm peak}/f_{\rm
  peak} (Q \rho/0.7) \sim 2.4 $.  We suspect this factor of $2.4$
comes from the fact that we are using different waveform templates for
parameter estimation.

Let us end this section by commenting on how the measurement accuracy
of the NS radius changes if the correct EOS in nature is not
TM1. Among the 5 EOSs considered in this paper, there is a very good
chance of detecting the post-merger signal after power stacking for
the DD2, TM1, LS220 and Shen EOSs, as discussed at the end of
Sec.~\ref{sec:stacking-vs-single}. The Shen EOS is quite similar to
TM1, so let us consider LS220 here. The approximation in
Eq.~\eqref{eq:delta-f} shows that $\delta
f_\mathrm{peak}/f_\mathrm{peak}$ depends only on $Q$ and $\rho$. From
Table~\ref{table:para} and comparing Figs.~\ref{fig:plot1}
and~\ref{fig:plot3}, one sees that $Q \, \rho$ for the DD2 EOS is
roughly a factor of $2$ smaller than $Q \, \rho$ for the TM1 EOS,
which leads to a $\delta f_\mathrm{peak}$ that is approximately a
factor of two larger, given the difference in
$f_\mathrm{peak}$. Furthermore, using Eq.~\eqref{eq:deltaR} one finds
that DD2 has a $\delta R_{1.6 M_\odot}$ that is roughly a factor of
$1.1$ times smaller. This means that if the post-merger signal is
detected with CE via power stacking, depending on the underlying EOS
we expect that $\delta f_\mathrm{peak}$ and $\delta R_{1.6 M_\odot}$
to lie in the range $\sim 4-20$ Hz and $15-56$m, respectively ($\sim
4\%$ accuracy). Thus, systematic errors seem to always dominate
statistical errors on the NS radius measurement irrespective of the
EOS for third generation detectors (for stacked signals that pass the
detection threshold).

\section{Discussion and Conclusion}\label{sec5}

In this work we have studied the possibility of detecting the GWs
generated by the oscillations of hypermassive NSs formed following BNS
mergers with future ground-based GW detectors. Based on the latest
estimates of the BNS merger rate and fitting formulas for the
oscillation peak frequency from state-of-the-art BNS merger
simulations, we found that the chance of detecting such oscillations
from individual sources could be low even for third generation GW
detectors, depending on the EOS. However, we point out that
detectability of individual events could potentially improve if one
considers all components/peaks that arise in the post-merger waveform,
and not only the dominant peak, as we do here. Nevertheless, it is not
currently clear whether sub-leading modes can persist and contribute
substantially to the SNR or whether their frequencies might drift,
hence making their detection challenging.

In order to increase the detection rate of the dominant post-merger
component, we 
 multiplied the Bayes factor
of each event to derive the Bayes factor of combined events. We refer
to this approach as power stacking.  Such an analysis was used to
propose a test of General Relativity in~\cite{Agathos:2013upa}, to
probe the BH no-hair property in~\cite{PhysRevD.90.064009} and to
explore EOS properties in~\cite{Agathos:2015uaa}. We have shown that this method can significantly 
boost the statistical chance of detection (shown in Fig.~\ref{fig:alpha}) as compared
to single events. The stacked signal can also be used to distinguish between
different NS EOSs. We formulated a Bayesian  model
selection framework, and illustrated its application by comparing EOS
model TM1 vs DD2, assuming the former is the true EOS. In practice,
such a model selection method only suggests {\it relative} preference
between the two selected models, both of which do not have to be the
true EOS. Thus, the results of model selection should be combined with
the signal-to-noise level of the stacked signal, assuming different
EOSs, to obtain an overall sense of the ``true" EOS.

The power stacking methods can be naturally applied to other
post-merger oscillation modes both in isolation or in combination with
other modes. For instance, one could apply this to the $21$ mode, which can become
strong if a one-arm instability develops in the BNS remnant
\cite{Paschalidis:2015mla,East:2015vix,East:2016zvv,Lehner:2016wjg}.
In this case, there is a tight correlation between the frequencies of the
$21$ and $22$ modes which can be exploited to further enhance the
achievable stacked SNR. This method can also be used to stack other post-merger
GW templates, such as the Principal Component basis
developed in \cite{clark2016observing}.  In an even broader context,
this approach could also be exploited to help identify decaying modes
in cold atom data~\cite{PhysRevLett.94.170404,Cao:2010wa} and their
connection with possible BH  duals through holographic
arguments~\cite{Brewer:2015ipa,Bantilan:2016qos}.

If in the future the theoretical uncertainty in modelling gets down 
to a level that the initial phase of post-merger modes can be estimated
given the binary parameters (including EOS), we can make
use of this phase information to
coherently stack a set of post merger events to further boost the collective
SNR. We demonstrated that such coherent stacking could significantly
increase the detection probability of the BNS post-merger dominant
$22$ mode. 
We explicitly
showed that if we require the same level of statistical significance,
then coherent stacking is more efficient at increasing the Bayes
factor than the current way power stacking is performed (at least in
the Bayesian framework we adopt); related comparisons are shown in
Figs.~\ref{fig:alpha} and~\ref{fig:na}. It would be interesting to
find a Bayesian formulation that mimics the behavior of coherent
stacking for low-SNR events, for example by introducing different
weights for different events.

The main limitation of the coherent stacking method is that it
requires small phase uncertainty in constructing the coherently
stacked signal. If the phase uncertainties are large, the coherent part 
of the stacked signal (the second line of Eq.~\eqref{eqts}) will be reduced dramatically. 
This could be alleviated if the initial phase of the $22$
mode can accurately be estimated using the inferred parameters of the inspiral
together with numerical simulations of the merger event. 
Though producing full templates of the post-merger signals incorporating all
the correct microphysics may not be practical within the next few
years, it may not be unreasonable to expect that simulations can at
least provide an accurate prediction of the initial phase of the 22
mode, as this will be fixed within the first few ms
post-merger\footnote{The instantaneous 22 mode phase may drift with
  time due to non-linear effects, but that is beyond the scope of this
  model.}.

Another limitation of the framework used in this paper is that
one needs to assume all the parameters are known except for 
$A_c$ and $A_s$ (or the amplitude and the phase offset). It would 
be interesting to extend the framework further to the case with unknown
$f_\mathrm{peak}$ and $Q$. Then, one does not need to assume 
the underlying EOS a priori, and one can reformulate the Bayesian
hypothesis test problem by taking into account the prior distribution
of $f_\mathrm{peak}$ and $Q$.

When the post-merger SNR is above unity, the inspiral SNR will be
large and one can likely extract nuclear physics information from the
measurement of NS tidal deformations that occur in this phase.  Thus,
it would be interesting to study how the post-merger detection via
stacking helps in probing nuclear physics by further including the
inspiral measurement. Universal relations between the post-merger
oscillation peak frequency and the leading tidal parameter in the
inspiral waveform~\cite{Takami:2014tva,Bernuzzi:2015rla} may help in
addressing this question.  Alternatively, an independent measurement
of the tidal deformability and the post-merger peak frequency may
allow one to confirm such universal relations from observations. If
such relations are altered from the GR prediction in modified
theories of gravity, one can use such a measurement to probe
strong-field gravity.  A similar proposal was already made and
demonstrated regarding the universal relation between the tidal
deformability and moment of
inertia~\cite{Yagi:2013bca,Yagi:2013awa,Yagi:2016bkt}. Also, 
as mentioned, complementary information from electromagnetic observations
-- coupled with refined numerical studies to connect the behavior of
cold and finite temperature nuclear EOS --
could be exploited to inform suitable priors for the analysis
described here.

\acknowledgements The authors thank R. O'Shaughnessy and P. Romatschke 
for interesting discussions. F.P., V.P., H.Y. and
K.Y. acknowledge support from NSF grant PHY-1607449 and the Simons
Foundation. V.P. also acknowledges support from NASA grant NNX16AR67G
(Fermi). K.Y. also acknowledges support from JSPS Postdoctoral
Fellowships for Research Abroad.  N.Y. is supported by NSF CAREER Grant PHY- 1250636 and NASA
grant NNX16AB98G to Montana State University. L.L. is supported in part
by NSERC and CIFAR. Research at Perimeter Institute is supported through 
Industry Canada and by the  Province of Ontario
through the Ministry of Research \& Innovation.

\appendix

\section{Fitting the post-merger dominant gravitational-wave mode}
\label{appendix:fits}

\begin{figure}[tb]
\includegraphics[width=8.4cm]{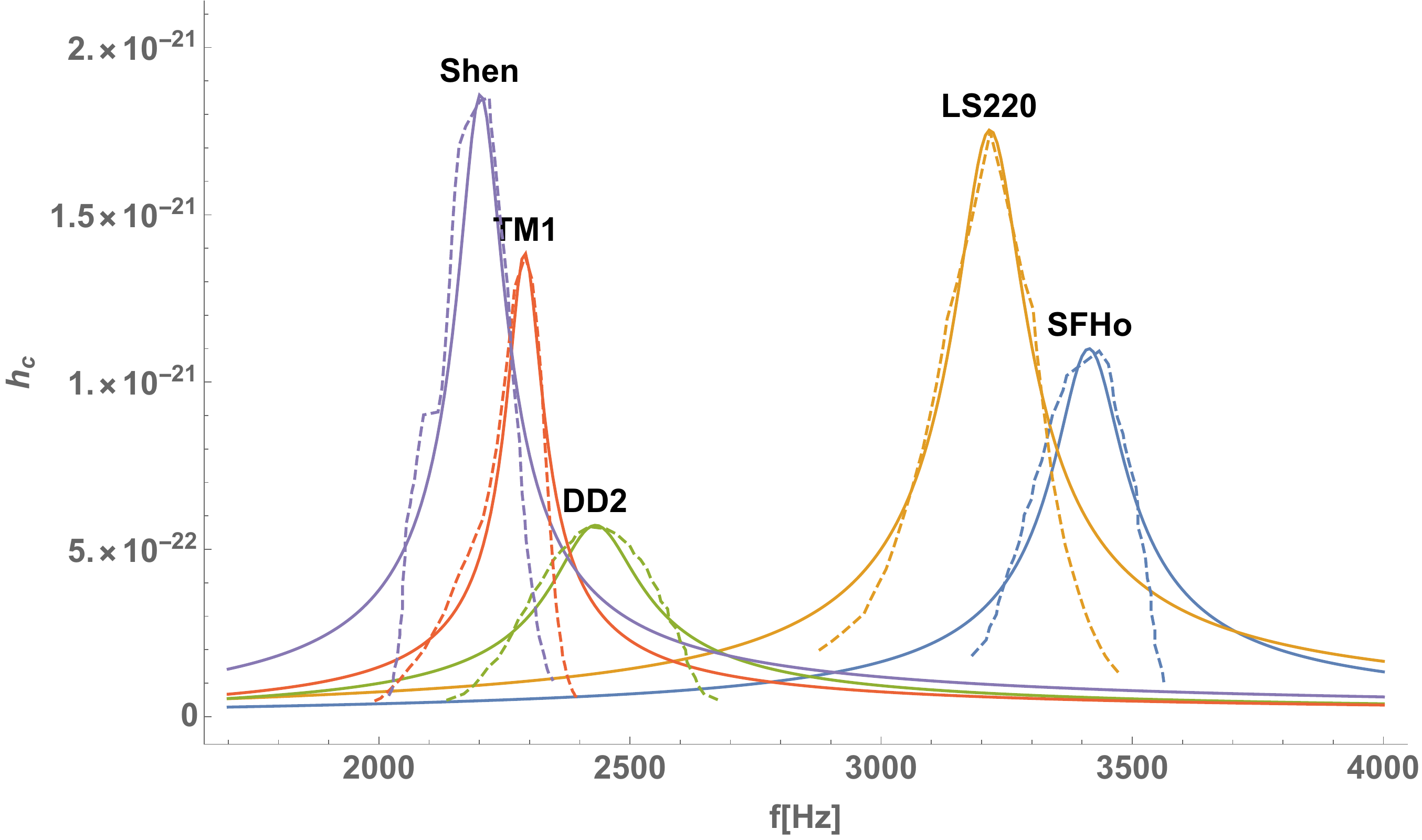}
\caption{Characteristic strain $h_c$ vs frequency for the dominant
  peak in BNS post-merger gravitational wave spectra and for various
  equations of state considered in this work. The luminosity distance
  to the source is set to 50 Mpc. Solid lines correspond
  to Eq.~\eqref{eq:w22} with the values for the amplitude and quality
  factor listed in Table~\ref{table:para}. Dashed lines correspond to
  the post-merger spectra obtained from numerical relativity
  simulations
  of~\cite{Sekiguchi:2011mc,Stergioulas:2011gd,Palenzuela:2015dqa}.}
\label{fig:hcfits}
\end{figure}

In this appendix, we show how well Eq.~\eqref{eq:w22} approximates the
dominant peak of the post-merger 22 mode obtained from the numerical
simulations
of~\cite{Sekiguchi:2011mc,Stergioulas:2011gd,Palenzuela:2015dqa}.  As
discussed in the main text the strain amplitude $A'$ and the quality
factor $Q$ in Eq.~\eqref{eq:w22} are chosen for each equation of state
such that the peak value of the characteristic strain and SNR with
Eq.~\eqref{eq:w22} match the peak value of the characteristic strain
and SNR of the dominant post-merger $22$ component found in the
corresponding BNS merger simulations. Figure~\ref{fig:hcfits} shows
that the Lorentzian profile~\eqref{eq:w22} provides a reasonable
approximation of numerical relativity post-merger spectra around the
dominant peak.

\section{Variance and Signal-to-noise Level of $\hat T$}
\label{appendixA}

In this appendix, we explain the variance and the signal-to-noise level of $\hat T$, 
which is the log of the Bayes factor between the 
hypotheses $\mathcal{H}_1$ and $\mathcal{H}_2$. Such a signal-to-noise level
can be used, instead of $\rho$, to discuss the detection criterion of the post-merger GW signals.

Let us begin with the single event case.
The variance of $n_T$ is given by 
\begin{align}
{\rm Var}[n_T]= &1+ A^2_c+A^2_s\,\nonumber \\
=& 1 +  \rho^2\,,
\end{align}
(recall that we chose $\langle c | c \rangle = 1 = \langle s | s \rangle$).
We can then define the ratio
\begin{align}
\frac{2 s_T}{\sqrt{{\rm Var}(n_T)}} = \frac{\rho^2}{\sqrt{1+ \rho^2}}\,,
\end{align}
which intuitively measures the signal-to-noise level in the GLRT
variable $\hat{T}_{\rm single}$. In the limit that the detection ${\rm
  SNR} = \rho \gg 1$, it is straightforward to see that the above
ratio is approximately $\rho$.

We now explain the coherent stacking case.
The noise part of $\hat{T}_{\rm coherent}$, i.e., ${\bf n}_{Ty}$,
follows a distribution similar to Eq.~\eqref{eqcomd}, with $A_c
\rightarrow \langle g_y |c \rangle$ and $A_s \rightarrow \langle g_y | s
\rangle $. Its variance is given by
\begin{align}\label{eqtn}
{\rm Var}[{\bf n}_{Ty}] 
= &  \left (\sum_i w^2_i \int df \frac{4 S_{n_i}(f) |\tilde h_c(f)|^2}{S_{n_y}(f)^2} \right )^2 \nonumber \\
+&  \sum_k w^2_k \sum_{i j } w_i w_j \int df \frac{2(\tilde g_i \tilde g^*_j e^{i ( \phi^0_i-\phi^0_j)}+c.c.) S_{n_k}}{S_{n_y}(f)^2}\,.
\end{align}
With Eq.~\eqref{eqts} and Eq.~\eqref{eqtn}, one can compute $\langle
{\bf s}_{Ty} \rangle/\sqrt{{\rm Var}[{\bf n}_{Ty}]}$ for the stacked signal
${\bf y}$. As before, this quantity is a measure of the signal-to-noise
level in the variable $\hat{T}_{\rm coherent}$.  Indeed, if all
individual events have the same SNR, then in the ${\rm SNR} = \rho \gg
1$ limit one finds $\langle {\bf s}_{Ty} \rangle/\sqrt{{\rm
    Var}[{\bf n}_{Ty}]} \approx N^{1/2} \rho$.

\section{Model  selection with different data sets}\label{appendixB}

For different EOSs we generically obtain different stacked signals,
thus one is faced with the problem of performing model selection using
different data sets as discussed in Sec.~\ref{sec4a}. In this appendix, we explain
how one can construct appropriate Bayes factors for such a model selection study.

We begin by generalizing further 
Eq.~\eqref{eq:BF} to allow different data sets:
\begin{align}\label{eqodd2}
\mathcal B_{12} \equiv \frac{P({\bf y}_1| \mathcal{H}_1)}{P({\bf y}_2| \mathcal{H}_2)}\,,
\end{align}
where ${\bf y}_1$ (${\bf y}_2$) is the data {\bf y} stacked using the
frequency scaling of EOS 1 (2).  Within the GLRT framework, one may
consider the expectation value and noise distribution of the random
variable
\begin{align}
\hat{\mathcal T}_{12} \equiv \log \mathcal B_{12}\,
\end{align}
to do model selection using the Jeffreys criteria. Notice that $\hat{\mathcal T}_{12} =-\hat{\mathcal T}_{21}$.  
Let us assume that model $1$ represents the true underlying EOS.  One
interesting feature implied by Eq.~\eqref{eq:fp} is that even if we
make an assumption that the EOS is model 2 where the true underlying
EOS follows model 1, the frequency rescaling factors depend only on
the total mass for each event, and of course the measured mass is EOS
independent.  Therefore, the main consequence of assuming an
``incorrect'' EOS is that the data from different events are not
coherently stacked onto each other due to the frequency mismatch
between the predicted signal and the actual signal.  Such a mismatch
also brings systematic errors on the phase measurement, making the
signals further incoherent. Incomplete coherent stacking may greatly
degrade the signal part of GLRT variable.

Alternatively, if model 2 is so incorrect that the $\rho$ of the
stacked signal is well below that of model 1, the phase
error in constructing the stacked signal is large and the weights
obtained by assuming an ``incorrect" EOS are far from their optimal
values, then ${\bf y}_1$ seems to be a convincingly better set of data than
${\bf y}_2$. Therefore, it is more appropriate to evaluate the following
Bayes factor:
\begin{align}\label{eqodd3}
\mathcal B_{1|2} \equiv \frac{P({\bf y}_1 | \mathcal{H}_1)}{P({\bf y}_1 | \mathcal{H}_2)}\,.
\end{align}
Intuitively, $\mathcal B_{12}$ may work better at distinguishing two
close EOSs, whereas $\mathcal B_{1|2}$ is expected to have wider
applicability and the associated analysis is more straightforward. In
this paper, we chose to study the statistical behavior of $\mathcal
B_{1|2}$, with model 1 being the one with better SNR from the
study in Sec.~\ref{sec31}. The extension of the analysis presented
here to the case dealing with the random variable $\mathcal B_{12}$
goes beyond the scope of the current paper.

\bibliography{master}
\end{document}